\documentclass[12pt,a4paper]{article}
\newlength{\overeqskip}
\newlength{\undereqskip}
\usepackage{amssymb}
\usepackage[dvips]{graphicx}
\usepackage[dvips]{color}
\def\nc{\newcommand}
\def\phm{\phantom{-}}
\nc{\half}{\frac{1}{2}}
\nc{\shalf}{\ensuremath{\textstyle \frac{1}{2}}}

\nc{\deldag}{\mathbin{\partial\mkern-10.5mu\big/}}
\nc{\deldagss}{\mathbin{\partial\mkern-10.5mu/}}
\nc{\kdag}{\mathbin{k\mkern-10mu\big/}}
\nc{\udag}{\mathbin{u\mkern-10mu\big/}}
\nc{\kdagss}{\mathbin{k\mkern-10mu/}}
\nc{\Pdag}{\mathbin{P\mkern-10mu\big/}}
\nc{\pp}{{\scriptscriptstyle ||}}
\nc{\stwo}{{\scriptscriptstyle 2}}
\nc{\pham}{{\phantom{-}}}

\def\lsim{\mathrel{\raise.3ex\hbox{$<$\kern-.75em\lower1ex\hbox{$\sim$}}}}
\def\gsim{\mathrel{\raise.3ex\hbox{$>$\kern-.75em\lower1ex\hbox{$\sim$}}}}
\def\apriori{{\em a priori }}

\def\Slashnew#1{#1\kern-0.55em\raise.05ex\hbox{/}}
\def\slashnew#1{#1\kern-0.5em\raise.05ex\hbox{{$\scriptstyle /$}}}
\def\emph#1{{\em #1}}
\def\hepph#1{hep-ph/#1}
\def\hepth#1{hep-th/#1}

\nc{\beq} {\begin{equation}}
\nc{\eeq} {\end{equation}}
\nc{\beqa}{\begin{eqnarray}}
\nc{\eeqa}{\end{eqnarray}}

\begin{document}
%
%
\begin{titlepage}
\pagestyle{empty}
\baselineskip=21pt
\vskip .6in

\begin{center} {\Large{\bf Quantum kinetic theory for fermions in temporally varying backrounds}}

\end{center}
\vskip .3in

\begin{center}

Matti Herranen,  Kimmo Kainulainen and Pyry Matti Rahkila\\

\vskip .3in

{\it  University of Jyv\"askyl\"a, Department of Physics, P.O.~Box 35 (YFL),
        \\ FIN-40014 University of Jyv\"askyl\"a, Finland }\\
\vskip .1in
and \\
\vskip .1in
{\it  Helsinki Institute of Physics, P.O.~Box 64, FIN-00014 University of  		
   	       Helsinki, Finland.}\\

\vskip .2in
\end{center}

\vskip 0.3in

\centerline{ {\bf Abstract} }
\baselineskip=16pt
\noindent
\\
We derive quantum kinetic equations for fermions in a homogeneous time-dependent background in presence of decohering collisions, by use of the Schwinger-Keldysh CTP-formalism. The quantum coherence (between particles and antiparticles) is found to arise from new spectral solutions
for the dynamical 2-point correlation function in the mean field limit. The physical density matrix $\rho$ and its dynamics is shown to be necessarily dependent on the extrenous information on the system, and expressions that relate $\rho$ to fundamental coherence functions and fermionic particle and antiparticle numbers are derived. For an interacting system we demonstrate how smooth decoherence effects are induced by collisions. As special applications we study the production of unstable particles during the preheating stage of the inflation and an evolution of an initially quantum $\rho$ towards a statistical limit including decoherence and thermalisation.

\vskip 2truecm
\noindent matherr@phys.jyu.fi, kainulai@phys.jyu.fi, pmrahkil@phys.jyu.fi

\end{titlepage}

\baselineskip=17pt

%
%

\section{Introduction}

We study the quantum dynamics of fermions in a homogeneous but temporally varying background field including collisions with a thermal background. These conditions are appropriate to model for example the particle production during preheating at the end of inflation or during cosmological phase tranistions~\cite{brandenbeger,Prokopec_partnumber}, as well as baryogenesis during preheating~\cite{shapo}, or coherent baryogenesis~\cite{cohbaryog}. They are relevant also for neutrino oscillations in the early universe~\cite{EKT}, or just for generic studies of thermalisation of quantum systems~\cite{generic_therm}. The formalism we will be developing here can also be modified for a treatment of static problems with planar symmetry~\cite{HKR1,HKRfuture}. In this form the resulting kinetic equations will be of interest for problems involving quantum reflection such as electroweak baryogenesis~\cite{CKV,CJK}.  For related studies of quantum transport equations for electroweak baryogenesis starting from similar theoretical foundation see {\em e.g.}~\cite{KPSW,PSW}.

By use of the CTP-formalism~\cite{Schwinger-Keldysh,CTP} we will set up quantum kinetic equations for the fermionic 2-point function $G^<$. We find that these equations admit a rich structure of spectral solutions including the expected mass-shell solutions for particle and antiparticle excitations, but also a new class of solutions living in $k_0=0$-shell in the phase space~\cite{HKR1}. We interpret  the arbitrary weight functions on these shells as describing the out-of-equilibrium particle and antiparticle numbers (mass shells) and the quantum coherence between particles and antiparticles ($k_0=0$-shell). New coherence solutions are shown to be excluded from the spectral function by the spectral sum-rule, which indicates that they are not part of the kinematic phase space, although they necessarily occur in the dynamical function $G^<$.  These solutions are also eliminated from $G^<$ by the KMS-conditions in the thermal limit. As a consequence of the singular shell structure, an integration procedure is needed in order to define a physical density matrix in terms of the original 2-point function. We show how this procedure necessarily involves specifying precisely the amount of information on the system, and derive an evolution equation for a density matrix relevant for a spatially homogenous case, including quantum coherence. We then introduce the interactions and show that the spectral structure for the phase space including the coherence shell solutions survives in the quasiparticle limit with the interactions.  We set up dynamical equations for the physical density matrix including the interactions and compute explicit expressions for the collision integrals in the case of a simple model Lagrangian describing decays and inverse decays.  The interaction terms are shown to contain the usual collision terms that push the mass-shell functions towards thermal limit, but also other collision terms that tend to bring the quantum coherence functions to zero. The loss of coherence does not happen instantly, like a collapse of a wave function, but smoothly over a characteristic time scale set by the strength of the interactions at the shell $k_0=0$, in close analogy to the damping of coherence in the case of neutrino oscillations~\cite{EKT}.   

In our approach we {\em define} the fermionic number density in the same way as in thermal field theory, that is, as an dimensionless real-valued function living on the positive mass-shell of the spectral form 2-point correlator. With this definition we see that in thermal limit the number density is indeed the standard Fermi-Dirac distribution. We will show that our definition for the particle number agrees with that of ref.~\cite{Prokopec_partnumber}, where it was derived using the operator methods and Bogolyubov transformation to diagonalize the fermionic Hamiltonian. We also derive expressions for the energy density and the pressure. The latter is shown to differ from the statistical pressure, but the statistical pressure is retrieved for any realistic measurement that averages out the quantum oscillations. As applications of our formalism we calculate particle number production during fermionic preheating including finite decay width for the heavy particles produced. We will see that the decoherence induced by the decays can have dramatic effect on the particle number evolution. We also show explicitly how an initially highly correlated out-of-equilibrium density matrix relaxes to a thermal equilibrium as a result of collisions.
  
This paper is organized as follows: In section \ref{sec:formalism} we give a brief intoduction to Shwinger-Keldysh CTP-formalism in order to write down the fermionic Kadanoff-Baym (KB) transport equations for 2-point correlation functions. In section \ref{sec:free} we study the noninteracting KB-(or Dirac) equation and find out the nontrivial phase-space shell structure including the new $k_0=0$-shell. In section \ref{sec:weighted} we define a nonsingular
physical density matrix, as a weighted integral over the initial singular 2-point function and show that its evolution is heavily dependent on the extrenous information we have on the system. The material in the sections \ref{sec:free}-\ref{sec:weighted} partly overlaps with the derivation in the companion paper~\cite{HKR1}, but we include a shortened discussion here for completeness. In section \ref{sec:physical} we will compute the particle number density, energy density and pressure in terms of the density matrix. In section \ref{sec:interactions} we generalize our kinetic equations to include collisions. We study the particle production at the preheating and the decoherence phenomenon in sections \ref{sec:preheating} and \ref{sec:decoherence}, and finally section \ref{sec:discussion} contains our conclusions. 

\section{General fermionic CTP-formalism}
\label{sec:formalism}
\begin{figure}
\centering
\includegraphics[width=0.75\textwidth]{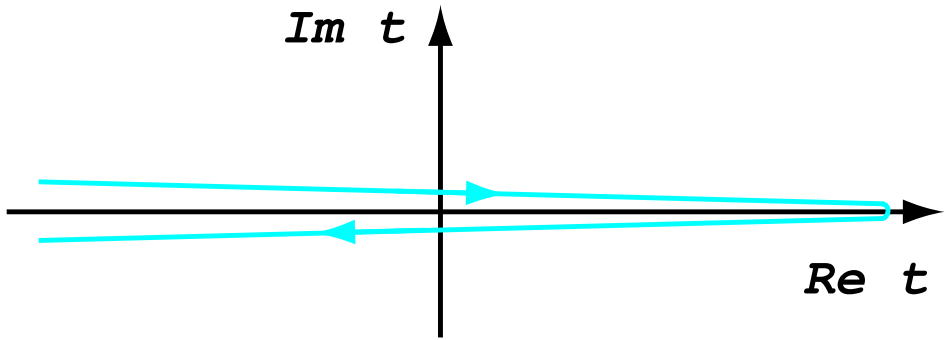}
     \caption{Schwinger-Keldysh path in complex time.}
     \label{fig:KeldyshPath}
\end{figure}

\begin{figure}
\centering
\includegraphics[width=0.75\textwidth]{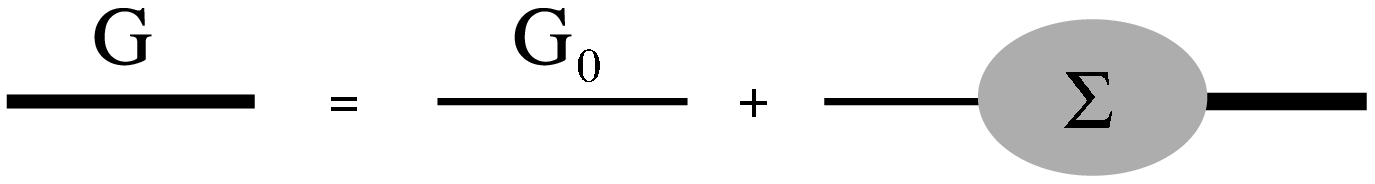}
     \caption{A generic form of a Schwinger-Keldysh equation for the
     2-point function $G^<$.}
     \label{fig:SchKelGen}
\end{figure}

The basic object of interest in this paper is the fermionic
2-point Wightmann function defined as:
\beq
  iG^<_{\alpha\beta}(u,v) \equiv
     \langle  \bar \psi_{\beta}(v){\psi}_{\alpha}(u)\rangle \equiv
     {\rm Tr}\left\{\hat \rho \ \bar \psi_{\beta}(v){\psi}_{\alpha}(u)\right\},
\label{G-less}
\eeq
where $\hat \rho$ is some unknown quantum density operator describing the properties of the system. In a non-interacting theory $G^<$ decouples from other $n$-point functions and the dynamical equation it satisfies is equivalent to the ordinary  Dirac equation. This problem was studied carefully in ref.~\cite{HKR1}. However, here we wish to include interactions, and so it is necessary to work in the framework of the quantum field theory. The QFT formalism that is well suited for the study of "in-in''-correlators\footnote{With "in-in''-correlator we mean that the matrix elements are expectation values $\langle in|A|in \rangle$, in contrast to the traditional QFT transition amplitudes $\langle out|A|in \rangle$ with different in and out states} like (\ref{G-less}) in possibly out-of-equilibrium conditions is called Schwinger-Keldysh or Closed Time Path (CTP) formalism~\cite{Schwinger-Keldysh,CTP}. In that formalism one defines a path ordered 2-point function on a complex Keldysh time-path (Dirac indices are suppressed): 
\beq
  iG_{\cal C}(u,v) =
            \left\langle T_{\cal C}
                 \left[\psi(u) \bar \psi (v)\right]
             \right\rangle ,
\label{Gcontour}
\eeq
where $T_{\cal C}$ defines time ordering along the Keldysh contour
${\cal C}$, which starts at some $t_0$, often taken to be at $-\infty$,
goes to $+\infty$, and then back to $t_0$ (see Fig.~\ref{fig:KeldyshPath}).
It can be shown for example by using the two-particle-irreducible (2PI)
effective action techniques \cite{2PI,CTP} that the 2-point function
$G_{\cal C}(x,y)$ obeys the contour Schwinger-Dyson equation:
\beq
G_{\cal C} (u,v) = G^0_{\cal C} (u,v)
             + \int_{\cal C} {\rm d}^4z_1 \int_{\cal C}
                             {\rm d}^4z_2 \; G^0_{\cal C} (u,z_1)
               \Sigma_{\cal C} (z_1,z_2) G_{\cal C} (z_2,v)\,,
\label{SD1}
\eeq
Equation (\ref{SD1}) is formally expressed in Fig.~\ref{fig:SchKelGen},
where the thin lines correspond to the free particle (tree level) propagator
$G^0_{\cal C}$, and the thick lines to the full propagator $G_{\cal C}$.
The filled ellipsis represents the self-energy function $\Sigma_{\cal C}$,
whose precise form depends on the model Lagrangian and a truncation scheme.
$\Sigma_{\cal C}$ couples $G_{\cal C}$ to an infinite (BBGKY-) hierarchy of
equations for higher (up to infinite) order Green's functions, and some
approximation scheme is needed to truncate this hierarchy to obtain the
closure. In the weak coupling limit it will be natural to do this by
substituting all higher than 2-point functions by their perturbative
expressions. However, we can learn a lot about the structure of the
SD-equations (\ref{SD1}) without ever making any reference to the  
explicit form of $\Sigma$. Multiplying Eq.~(\ref{SD1}) by the inverse of the free particle propagator $(G^0_{\cal C})^{-1}$ and integrating over the connecting variable  $z_1$ one finds
\beq
   \int_{\cal C} {\rm d}^4z G^0_{\cal C}(u,z)^{-1} G_{\cal C}(z,v)
                =  \delta_{\cal C}(u-v)
                + \int_{\cal C} {\rm d}^4z
                     \Sigma_{\cal C}(u,z) G_{\cal C}(z,v),
\label{SD2}
\eeq
where $\delta_{\cal C}(u-v) \equiv \delta_{\cal C}(u^0_{\cal C}-v^0_{\cal C})
\delta^3(\vec u - \vec v)$ is a contour time delta-function.
The complex time Green's function in (\ref{Gcontour}-\ref{SD2}) can be  
conveniently decomposed in four different 2-point functions with respect to
usual real time variable:  
\def\phm{\phantom -}
\begin{eqnarray}
  iG^<(u,v) \equiv -iG^{+-}(u,v)      &\equiv& \langle \bar\psi(v)\psi(u) \rangle
   \nonumber\\
  iG^>(u,v) \equiv \phm iG^{-+}(u,v)      &\equiv& \langle \psi(u)\bar\psi(v) \rangle
   \nonumber\\
  iG_F(u,v) \equiv \phm iG^{++}(u,v)      &\equiv&   \theta(u_0-v_0) G^>(u,v)
                                               - \theta(v_0-u_0) G^<(u,v)
   \nonumber\\
  iG_{\bar F}(u,v) \equiv \phm iG^{--}(u,v) &\equiv&  \theta(v_0-u_0) G^>(u,v)
                                               - \theta(u_0-v_0)G^<(u,v)\,.
\label{GFs}
\end{eqnarray}
where $G_F$ and $G_{\bar F}$ are the chronological (Feynman) and
anti-chronological (anti-Feynman) Green's functions, respectively, and $G^<$
and $G^>$ are the (quantum) Wightmann distribution functions. 
Similar
decomposition can be done for the contour self-energy $\Sigma_{\cal C}$. By using the 2PI effective action techniques the  self-energies on different branches are obtained by the following functional differentiation (see eg.~\cite{PSW}):
\beq
\Sigma^{ab}(u,v) \equiv -iab \frac{\delta \Gamma_2[G]}{\delta
  G^{ba}(v,u)} \,,
\label{2PIsigmas}
\eeq
where $\Gamma_2$ is the sum of all two particle
irreducible vacuum graphs of the theory and the indices $a,b$
refer to the position of the arguments $u$ and $v$, respectively, on
the complex Keldysh time path (for example $a=+1(-1)$ implies that
$u$ belongs to the upper (lower) branch of the time contour in Fig.~\ref{fig:KeldyshPath}.) We use the same
notation $\Sigma^< = -\Sigma^{+-}$, etc. as for propagators
(\ref{GFs}). Using these definitions and the relations: $\int_{\cal C}
{\rm d}^4u \rightarrow \sum_a a\int_{-\infty}^\infty{\rm d}^4u$ and
$\delta_{\cal C}(u^0_{\cal C}-v^0_{\cal C}) \rightarrow
a\delta_{ab}\delta(u^0-v^0)$ between complex Keldysh time and the
usual real time variables, we can write Eq.~(\ref{SD2}) in the
following matrix form: 
\beq
G_0^{-1} \otimes G = \sigma_3 \; \delta +  \Sigma \otimes \sigma_3 G,
\label{SD3}
\eeq
where
\beq
G=\left(\begin{array}{cc}
            G_F & - G^< \\
            G^> & \phantom{-} G_{\bar F}
         \end{array}\right) \qquad , \qquad 
\Sigma=\left(\begin{array}{cc}
            \Sigma_F & - \Sigma^< \\
            \Sigma^> & \phantom{-} \Sigma_{\bar F} \,,
         \end{array}\right) 
\label{Gmatrix}
\eeq
and $\sigma_3$ is the usual Pauli matrix, and we defined a shorthand
notation $\otimes$ for the convolution integral:
\beq
    f \otimes g \equiv \int {\rm d}^4z f(u,z)g(z,v).
\label{otimes}
\eeq
We have also left out the labels $u$ and $v$ where obvious; for example
$\delta \equiv \delta^4(u-v)$.
\subsection{Kadanoff-Baym equations}
It's appropriate to further define the retarded and advanced propagators (a similar decomposition obviously holds for the self energy  function $\Sigma$):
\begin{eqnarray}
  G^r(u,v)  &\equiv&  \phantom{-} \theta(u^0-v^0) (G^< + G^>) \nonumber\\
  G^a(u,v)  &\equiv&  -\theta(v^0-u^0) (G^< + G^>).
\label{raGFs}
\end{eqnarray}
Equations (\ref{SD3}) take on a particularily compact from when  
written in terms of these new Green's functions:
\begin{eqnarray}
   (G_0^{-1}-\Sigma ^{r,a})\otimes G^{r,a} &=& \delta
\label{KB1a} \\
   (G_0^{-1}-\Sigma ^{r})  \otimes G^{<,>} &=& \Sigma ^{<,>}\otimes  
G^{a}.
\label{KB1b}
\end{eqnarray}
Equations (\ref{KB1a}) and (\ref{KB1b}) are called \textit{pole equations} and \textit{Kadanoff-Baym (KB) equations}, respectively. In general, the former will fix the spectral properties of the theory, while the latter will give the dynamical evolution, including quantum transport effects. Indeed, in the classical limit the KB-equations (\ref{KB1b}) will reduce to well known Boltzmann transport equation for the phase space number density \cite{Heinz,KPSW,PSW}. 

It can be easily shown that the defined 2-point functions have the following hermiticity properties: 
\begin{eqnarray}
  \left[iG^{<,>}(u,v)\gamma^0\right]^\dagger &=& iG^{<,>}(v,u)\gamma^0
  \nonumber\\
  \left[iG^{r}(u,v)\gamma^0\right]^\dagger   &=& -iG^{a}(v,u)\gamma^0.
\label{CEq1}
\end{eqnarray}
These identities suggest to decompose $G^{r,a}$ into Hermitian and
anti-Hermitian parts:  
\begin{eqnarray}
  G_H       &\equiv&  \frac{1}{2}\left(G^a + G^r\right) \nonumber\\ 
  {\cal A}  &\equiv&  \frac{1}{2i}\left(G^a - G^r\right) 
                     = \frac{i}{2}\left(G^< + G^>\right).  
\label{Hdecomposition}
\end{eqnarray}
The anti-Hermitian part ${\cal A}$ is called the \textit{spectral function}.
Based on (\ref{raGFs}) it is easy to show that $G_H$ and ${\cal
  A}$ obey the spectral relation:
\beq
G_H(u,v) = -i {\rm sgn}(u^0-v^0) {\cal A} (u,v).
\label{spectrel}
\eeq
Since the self-energies $\Sigma$ satisfy identities similar to
(\ref{CEq1}), it's appropriate to define the Hermitian and
anti-Hermitian parts of $\Sigma^{r,a}$ as well:
\begin{eqnarray}
  \Sigma_H  &\equiv&  \frac{1}{2}\left(\Sigma^a + \Sigma^r\right) \nonumber\\ 
  \Gamma    &\equiv&  \frac{1}{2i}\left(\Sigma^a - \Sigma^r\right) 
                     = \frac{i}{2}\left(\Sigma^< + \Sigma^>\right)\,.  
\label{gamma}
\end{eqnarray}
By using the definitions (\ref{Hdecomposition}) and (\ref{gamma}) it is now
straightforward to show that the pole equations (\ref{KB1a}) can be
written in the form: 
\begin{eqnarray}
   (G_0^{-1}-\Sigma_H) \otimes {\cal A} - \Gamma
   \otimes G_H = 0
\nonumber \\
    (G_0^{-1}-\Sigma_H) \otimes G_H + \Gamma
   \otimes {\cal A} = \delta \,.
\label{SpecEq1}
\end{eqnarray}
while the KB-equation (\ref{KB1b}) for $G^<$ gives:
\beq
   (G_0^{-1}-\Sigma_H) \otimes G^< - \Sigma^< \otimes G_H
   = \frac{1}{2}\left( \Sigma^> \otimes G^< - \Sigma^< \otimes G^>  
\right)
\label{Dyneq}
\eeq
Assuming we will solve the spectral function from the pole equations
(\ref{SpecEq1}), we don't need to consider the other KB-equation for
$G^>$, since from the definition (\ref{Hdecomposition}) it immediately
follows that
\beq
G^> = - G^< - 2i {\cal A} \,.
\label{Gglrel}
\eeq
Further, we know that the canonical equal time anticommutation relation of the field operators
\beq
\{\psi(t,\vec u), \psi^\dagger (t,\vec v)\}  =  -i \delta^3(\vec u- 
\vec v)
\label{equalcomm}
\eeq
must be satisfied by all physical field configurations. 
Using definitions (\ref{GFs}) and (\ref{Hdecomposition}) it is easy to see that relations (\ref{equalcomm}) imply the condition
\beq
2 {\cal A}(t,\vec u; t, \vec v) \gamma^0 =  \delta^3(\vec u-\vec v)
\label{spectrel1}
\eeq
on the spectral function. This is the direct space version of the spectral sum-rule. It follows also directly from Eqs.~(\ref{spectrel}) and (\ref{SpecEq1}) without a reference to the commutation relation (\ref{equalcomm}). Note that apart from a possible implicit dependence of $\Gamma$ on $G^{<,>}$ the pole equations (\ref{SpecEq1}) are entirely independent of dynamical evolution. This is exactly as it should be; the dynamical evolution can affect the spectral solutions related to the kinematic phase space  only indirectly by changing the ambient conditions in the plasma in which the particles are moving.
\subsection{Lagrangian density and the mixed representation}
We now want to write the spectral and dynamical equations in detail in the mixed representation, by Fourier transforming the 2-point functions with respect to the relative coordinate $r \equiv u-v$. This representation is useful in dealing with the evolution equations because it performs a separation of the internal and external distance scales of the problem, and easily allows expansions in the gradients in a (slowly varying) external coordinate. At this point we also wish to specify explicitly the free particle Green's function $G_0$; in this paper we will consider the following CP-violating fermionic Lagrangian
\beq
{\cal L} = i\bar \psi \deldag \psi
                    + \bar \psi_L m \psi_R
                    + \bar \psi_R m^* \psi_L + {\cal L}_{\rm
                      int} \,,
\label{freeLag1}
\eeq
where $m(x) = m_R(x) + im_I(x)$ is complex, possibly spacetime
dependent mass and ${\cal L}_{\rm int}$ is the interaction part to be
defined later. From Eq.~(\ref{freeLag1}) it's easy to see that 
\beq
G_0^{-1}(u,v) = \delta^4(u-v) (i \deldag_v - m^*(v) P_L - m(v)  
P_R)\,,
\label{free-prop}
\eeq
where $P_{L,R} = \frac{1}{2}(1 \mp \gamma^5)$. Next, we define the Wigner transformation of an arbitrary 2-point function as follows:
\beq
F(k,x) \equiv \int d^{\,4} r \, e^{ik\cdot r}
                                         F(x + r/2,x-r/2).
\label{wigner1}
\eeq
where $x$ is the average coordinate, and $k$ is the internal momentum
variable conjugate to relative coordinate $r$. Performing the Wigner
transformation to Eqs.~(\ref{SpecEq1}) and (\ref{Dyneq}) we get the 
pole- and KB-equations in the mixed representation:
\begin{eqnarray}
  (\kdag + \frac{i}{2} \deldag_x - \hat m_0
             - i\hat m_5 \gamma^5) {\cal A}
  -  e^{-i\Diamond}\{ \Sigma_H \}\{ {\cal A}\}
  -  e^{-i\Diamond}\{ \Gamma \}\{G_H\} &=& 0
\label{SpecEqMix1} \\
  (\kdag + \frac{i}{2} \deldag_x - \hat m_0
             - i\hat m_5 \gamma^5) G_H
  -  e^{-i\Diamond}\{ \Sigma_H \}\{ G_H \}
  +  e^{-i\Diamond}\{ \Gamma \}\{{\cal A}\} &=& 1
\label{SpecEqMix2}
\end{eqnarray}
and
\beq
(\kdag + \frac{i}{2} \deldag_x - \hat m_0
       - i\hat m_5 \gamma^5) G^<
  -  e^{-i\Diamond}\{ \Sigma_H \}\{ G^< \}
  -  e^{-i\Diamond}\{ \Sigma^< \}\{ G_H \}
  = {\cal C}_{\rm coll},
\label{DynEqMix}
\eeq
where the collision term is given by
\beq
{\cal C}_{\rm coll} \equiv \frac{1}{2} e^{-i\Diamond}
                             \left( \{\Sigma^>\}\{G^<\} - \{\Sigma^<\}\{G^>\}
                             \right) \,.
\label{collintegral}
\eeq
The $\Diamond$-operator is the following generalization of the Poisson brackets:
\beq
\Diamond\{f\}\{g\} = \frac{1}{2}\left[
                   \partial_X f \cdot \partial_k g
                 - \partial_k f \cdot \partial_X g \right]
\label{diamond}
\eeq
and the mass operators $\hat m_0$ and $\hat m_5$ are defined as:
\beq
\hat m_{\rm 0,5} F(k,x) \equiv m_{R,I}(x) e^{-\frac{i}{2}
       \partial_x^m \cdot \partial_k^F} F(k,x)\,.
\label{massoperators}
\eeq
Transforming Eq.~(\ref{spectrel1}) in the same way gives the well known momentum space representation of the spectral sum-rule: 
\beq
  \int \frac{{\rm d}k_0}{\pi}
                  {\cal A}(k,x) \gamma^0  = 1.
\label{sumrule2}
\eeq
Equations (\ref{SpecEqMix1}-\ref{DynEqMix}) together with the sum-rule (\ref{sumrule2}) and the identity (\ref{Gglrel}) form a complete set of equations for solving $G^<$ and the pole functions ${\cal A}$ and $G_H$ exactly, when the interactions (a scheme to compute $\Sigma$) and the mass profiles are specified. In practice these equations are too hard to be solved in their full generality, and sevaral approximations are needed to find a solvable set of equations. It is not clear \apriori that a tractable approximation scheme that is general enough to treat information on quantum coherence simultaneously with interactions can be found. The novelty of this work is to show that such a scheme indeed does exist. We shall now proceed to build this scheme by first constructing the full phase space structure of the free 2-point functions.

\section{Free fields}
\label{sec:free}

In the noninteracting case the equation (\ref{DynEqMix}) for the dynamical function $G^<$ decouples from the pole functions. In this case, the Hermitian Wightmann function, defined as
\beq
\bar G^<(u,v) \equiv i G^<(u,v)\gamma^0.
\label{herm_gless}
\eeq
obeys the free Kadanoff-Baym equation in the mixed representation: 
\beq
\Big( k_0 + \frac{i}{2}\partial_t
    - \vec \alpha \cdot (\vec k - \frac{i}{2}\vec \nabla)
    - \gamma^0 \hat m_0 - i\gamma^0 \gamma^5 \hat m_5
\Big) \bar G^<(k,x) = 0 \,,
\label{G-lessEq3}
\eeq
which is obtained from Eq.~(\ref{DynEqMix}) by setting $\Sigma^{ab} = 0$ and multiplying from both sides by $\gamma^0$. In a spatially homogenous case the spatial gradient terms vanish and, correspondingly, the helicity is a good quantum number. Mathematically this follows from the fact that the helicity operator $\hat h= \hat k \cdot \vec S = \hat k \cdot \gamma^0 \vec \gamma \gamma^5$, where $\hat k \equiv \vec k/|\vec{k}|$, commutes with the differential operator of Eq.~(\ref{G-lessEq3}) in the homogenous limit. This fact is particularily transparent in Weyl basis where the gamma-matrices are given by the following direct product expressions:
\beq
\gamma^0 = \rho^1 \otimes 1 \,,
\qquad
\vec \alpha = -\rho^3 \otimes \vec \sigma
\quad {\rm and} \quad \gamma^5 = -\rho^3 \otimes 1 \,.
\label{gammamatrices}
\eeq
Here both $\rho^i$ and $\sigma^i$ are the usual Pauli matrices such that the
$\rho$-matrices refer to the chiral- and $\sigma$-matrices to the spin-degrees
of freedom.  In this representation the helicity operator is just $\hat h = 1
\otimes \hat k \cdot \vec \sigma$ and it's commutativity with $\vec
\alpha \cdot \vec k$, $\gamma^0$ and $\gamma^5$ is evident. As a
result one can introduce a decomposition of $\bar G^<$ in the helicity
basis: 
\beq
  \bar G_h^<  \equiv g_h^< \otimes
  \frac12(1 + h \hat k\cdot \vec \sigma),
\label{connectionHOMOG}
\eeq
where $g_h^<$ are unknown Hermitian $2\times 2$ matrices (for $h = \pm 1$) in chiral indices.  When this decomposition is inserted into Eq.~(\ref{G-lessEq3}) one obtains an equation
\beq
\Big( k_0 + \frac{i}{2}\partial_t \Big) g_h^< = \hat H  g_h^< \,,  
\qquad  {\rm with} \qquad 
\hat H \equiv  -h |\vec{k}| \rho^3 + {\hat m}_0 \rho^1 - {\hat m}_5 \rho^2 \,.
\label{Gs-eomHOMOG}
\eeq
Even this equation would be impossible to solve exactly, because the mass operators $ {\hat m}_{0,5}$ involve gradients to arbitrary orders. In the {\em mean field limit} the gradients drop out however, and $\hat H$ becomes a local Hermitian Hamilton operator
\beqa
\hat H \rightarrow - h |\vec{k}| \rho^3 + m_R \rho^1 + m_I \rho^2
= \left( \begin{array}{cc}
                   -h|\vec{k}| & m \\
                   m^* & h|\vec{k}|
                 \end{array}\right)  \equiv H\,.
\label{Hamilton}
\eeqa
One can always decompose the equation (\ref{Gs-eomHOMOG}) into two distinct sets of equations based on hermiticity. In the mean field limit, where $H=H^\dagger$, the {\em Hermitian} (H) and {\em anti-Hermitian} (AH) parts become:
\beq
{\rm (H)}:\quad 2 k_0 g_h^< = \{ H,g_h^< \}\,, \qquad
{\rm (AH)}:\quad i\partial_t g^<_h = [H, g^<_h] \,.
\label{Heq}
\eeq
respectively. Note that (H)-equation is an algebraic matrix equation because it does not contain any time derivatives. It will give rise to phase space constraints for the components of $g_h^<$, as will be shown in detail in the following subsection.  The (AH)-equation in (\ref{Heq}) on the other hand contains an explicit time derivative of $g_h^<$ and is thus called {\em ``kinetic equation}''.  This equation clearly has the standard form of the equation of motion of a density matrix in the Schr\"odinger picture. Indeed, it is easy to see that in the homogenous case the Dirac equation for the wave function $\psi = (L_h,R_h)$ (displaying only the nontrivial chiral components) of a given helicity $h$ becomes just
\beq
i\partial_t \psi = HÊ\psi \,,
\label{diraceq}
\eeq
where $H$ is given by the mean field limit Eq.~(\ref{Hamilton}). Given Eq.~(\ref{diraceq}) the equation of motion of the form (AH)-equation in (\ref{Heq}) follows immediately for $\rho \equiv \psi\psi^\dagger$. The Hamiltonian $H$ clearly has eigenvalues corresponding to free particle and antiparticle states: 
$k_0 = \pm \omega_k = \pm (\vec{k}^2 + |m|^2)^{1/2}$, and Eq.~(\ref{diraceq}) describes the mixing of these states in the case of a time dependent mass term. Despite the apparent similarities to the Dirac equation approach, equations (\ref{Heq}) are mathematically very different from Eq.~(\ref{diraceq}). Indeed, it turns out that the (H)-equations {\em impose} a singular shell structure for $g^<_h$ that prevents us from interpreting it directly as a physical density matrix and which consequently makes the (AH)-equation meaningless before a sensible integration procedure is defined.

\subsection{Shell structure}

The (H)-equation (\ref{Heq}) is most conveniently discussed in the Bloch-representation for $g_h^<$: 
\beq
g^<_h \equiv \frac12 \left( g^h_0 +  g^h_i \rho^i \right),
\label{glesss}
\eeq
where $g^h_\alpha$ are real-valued functions, because of the hermiticity of $g^<_h$. In the Bloch-representation (\ref{glesss}) the (H)-equation (\ref{Heq}) decomposes into the following four real-valued {\em``constraint equations''} (CE):
\beqa
k_0 g^h_0 + h|\vec{k}| g^h_3 - m_R g^h_1 + m_I g^h_2 &=& 0
\nonumber \\
k_0 g^h_3 + h|\vec{k}| g^h_0 &=& 0
\nonumber \\
k_0 g^h_1 - m_R g^h_0 &=& 0
\nonumber \\
k_0 g^h_2 + m_I g^h_0 &=& 0 \,.
\label{Hset1}
\eeqa
This set of linear homogeneos equations can be written as an equation $B_{\alpha\beta}g^h_\beta = 0$, where the coefficient matrix is 
(index ordering is here defined as $\alpha = 0,3,1,2$):
\beq
B = \left( \begin{array}{cccc}
    k_0  &  h|\vec{k}|  &  -m_R  &  m_I \\
    h|\vec{k}| &  k_0   &  0     &  0   \\
    -m_R &  0     &  k_0   &  0   \\
    m_I  &  0     &  0     &  k_0 
    \end{array} \right)
\eeq
A homogeneous matrix equation may have a nonzero solution only when the determinant of the matrix vanishes. Here the determinant is easily evaluated to give:
\beq
\det(B) = \left( k_0^2 - \vec{k}^2 - |m|^2 \right)k_0^2 \,,
\label{constraint_det}
\eeq
The vanishing of the determinant (\ref{constraint_det}) yields two different classes of spectral solutions with different dispersion relations. First, there are the usual mass-shell solutions with $k_0^2 - \vec{k}^2 - |m|^2 = 0$, but we find also new $k_0 = 0$-shell solutions. These solutions turn out to be the way the quantum coherence effects are introduced in the present approach. Let us next examine the matrix-stucture of these spectral solutions.

\subsubsection{$k_0 \neq 0$ -solutions; free particle mass-shell}

Let us first assume that $k_0 \neq 0$. Then the constraint equations (\ref{Hset1}) clearly have the solution:
\beq
  g^h_3 = - \frac{h|\vec{k}|}{k_0} g^h_0, \qquad g^h_1 = \frac{m_R}{k_0} g^h_0,
  \qquad g^h_2 = - \frac{m_I}{k_0} g^h_0,
\label{g13}
\eeq
and
\beq
 (k_0^2 - \vec{k}^2 - |m|^2) g^h_0 = 0.
\label{SpecEq}
\eeq
Equation (\ref{SpecEq}) has the spectral solution
\beqa
g^h_0(k_0,|\vec{k}|;t) &=& 4\pi \, f^h_{s_{k_0}}(|\vec{k}|,t) |k_0| \,\delta(k^2 - |m|^2) \nonumber \\
&=& 2\pi \, f^h_{s_{k_0}}(|\vec{k}|,t) \,\delta(k_0 - s_{k_0}
\sqrt{\vec{k}^2 + |m|^2}) \,,
\label{SpecSol}
\eeqa
where $s_{k_0} \equiv {\rm sgn}(k_0)$. So these solutions indeed live on
the energy-momentum mass-shell corresponding to the dispersion relation
\beq
k_0 = \pm \omega_k \equiv \pm \sqrt{\vec{k}^2 + |m|^2}.
\label{DR1}
\eeq
Now, using (\ref{g13}) and (\ref{SpecSol}) we can write the mass-shell
contribution for the full chiral $g^<_h$-matrix as follows:
\beq
g^<_{h,{\rm m-s}}(k_0,|\vec{k}|;t)  =
2 \pi f^h_{s_{k_0}}(|\vec{k}|,t) |k_0|
                                  \left(\begin{array}{cc}
                                  1 - h|\vec{k}|/k_0  &  m/k_0 \\
                                  m^*/k_0 &  1 + h|\vec{k}|/k_0
                                  \end{array} \right)
                                 \delta(k^2 - |m|^2).
\label{simpleqs}
\eeq
This solution clearly describes either a particle or an antiparticle eigenstate of helicity $h$ and momentum $\vec k$.

\subsubsection{$k_{0}=0$-solutions}

Setting $k_0=0$ in the first place we find out that equations (\ref{Hset1}) have a new class of solutions, which obey the relations
\begin{eqnarray}
g_3^h & = & h\frac{m_R}{|\vec{k}|} g_1^h - h\frac{m_I}{|\vec{k}|} g_2^h
\nonumber \\ 
g_0^h & = & 0,
\label{q-coh_HOMOG}
\end{eqnarray}
while the components $g_{1,2}^h$ are unconstrained. The corresponding spectral solution is then
\begin{eqnarray}
 g^<_{h,{\rm 0-s}}(k_0,|\vec{k}|;t) &=&  \pi \left[
                      f^h_{1}(|\vec{k}|,t) \left( \begin{array}{cc}
                                  h\, m_R/|\vec{k}| &  1 \\
                                  1       & - h\, m_R/|\vec{k}|
                                 \end{array} \right)
                             \right.
  \nonumber \\    && \phantom{m} + \left.
                   f^h_{2}(|\vec{k}|,t) \left( \begin{array}{cc}
                                  -h\, m_I/|\vec{k}| &  -i \\
                                  i       &  h\, m_I/|\vec{k}|
                                 \end{array} \right)
                    \right] \,
                    \delta (k_0) \,,
\label{k0zerospec}
\end{eqnarray}
where $f^h_{1}(|\vec{k}|,t)$ and $f^h_{2}(|\vec{k}|,t)$ are new unknown real functions. These solutions live on the $k_0 = 0$-shell, and they cannot be related to particles and antiparticles alone, since those should have energies $k_0 = \pm \omega_k$ respectively. On the other hand one would expect that the density matrix should somehow contain the information of the quantum coherence between the particles and antiparticles, just as does the corresponding Dirac equation. Thus we make a natural interpretation: \textit{the additional $k_0=0$-solutions (\ref{k0zerospec}) describe the quantum coherence between particles and antiparticles with same helicity $h$ and opposite momenta.} 

Combining the solutions (\ref{simpleqs}) and (\ref{k0zerospec}) gives the most complete solution that satisfies the constraint equations (\ref{Hset1}) for a given helicity $h$ and momentum $|\vec{k}|$:
\begin{eqnarray}
g^<_h(k_0, |\vec{k}|;t) &=& g^<_{h,{\rm m-s}}(k_0,|\vec{k}|;t) \, + \, g^<_{h,{\rm 0-s}}(k_0,|\vec{k}|;t)
\nonumber \\[3mm] 
  &=& 2 \pi f^h_{s_{k_0}}(|\vec{k}|,t) |k_0|
                                 \left( \begin{array}{cc}
                                  1 - h |\vec{k}|/k_0  &  m/k_0 \\
                                  m^*/k_0       &  1 + h |\vec{k}|/k_0
                                 \end{array} \right)
                                 \delta(k^2 - |m|^2).
\nonumber \\
  &+&  \pi \left[
                      f^h_{1}(|\vec{k}|,t) \left( \begin{array}{cc}
                                  h\, m_R/|\vec{k}| &  1 \\
                                  1       & - h\, m_R/|\vec{k}|
                                 \end{array} \right)
                             \right.
  \nonumber \\    && \phantom{m} + \left.
                   f^h_{2}(|\vec{k}|,t) \left( \begin{array}{cc}
                                  -h\, m_I/|\vec{k}| &  -i \\
                                  i       &  h\, m_I/|\vec{k}|
                                 \end{array} \right)
                    \right] \,
                    \delta (k_0) \,,
\label{fullchiralHOMOG}
\end{eqnarray}
This solution contains both particle and antiparticle states as well as their coherence in separate singular shells in the phase space, as we promised.  The solution (\ref{fullchiralHOMOG}) is a distribution which is best understood as a functional phase space measure, parametrized by weight functions that must be the real physical objects whose evolution we are interested in. It is clear that setting the form (\ref{fullchiralHOMOG}) directly into the (AH)-equation in (\ref{Heq}) does not lead into a sensible equation of motion however. Before discussing this problem further, we shall first discuss the spectral structure of the pole functions ${\cal A}$ and $G_H$.

\subsection{Spectral function and $G_H$}
\label{sec:spectral}

In the noninteracting case the equation of motion for the spectral function ${\cal A}$ Eq.~(\ref{SpecEqMix1}) is identical to that for $G^<$. As a result, the most general solution for ${\cal A}$ is of the form of Eq.~(\ref{connectionHOMOG}):
\beq 
  {\cal A} \gamma^0 = \sum_h a_h \otimes \frac12(1 + h \hat k\cdot
  \vec \sigma) \,, 
\label{specA1HOMOG} 
\eeq
where the chiral matrix $a_h$ is identical to the most general solution (\ref{fullchiralHOMOG}) for $g^<_h$, with four yet undefined spectral on-shell functions $f^{h{\cal A}}_\alpha$ for both helicities. However, the spectral function must in addition obey the sum-rule Eq.~(\ref{sumrule2}). This is enough to completely fix the values of all four on-shell functions:
\beq 
f^{h\,\cal A}_\pm = \frac12 \quad\quad {\textrm{and}} \quad\quad f^{h\,\cal
  A}_{1,2} = 0 \,. 
\label{specAconstrHOMOG}
\eeq
for both helicities. With these values the full solution for ${\cal A}$ becomes:
\beq
 {\cal A} = \pi {\rm sgn}(k_0) (\kdag +  m_R - i\gamma^5 m_I)
            \delta(k^2-|m|^2) \,.
\label{specA3}
\eeq
This is just the familiar result for the spectral function in thermal quasiparticle limit (see for example~\cite{PSW}). The spectral function is now completely determined, and it doesn't contain any dynamics at all. Especially, it does not have any contribution from the $k_0=0$-shell describing the coherence between particles and antiparticles. This is what we should expect, since coherence is a dynamic phenomenon, so it should not show up in the measure of the one-particle phase space. Moreover, it should be vanishing in the statistical equilibrium limit. 

For completeness we consider also the pole-function $G_H$ in the noninteracting case, although we shall not need this function later in the paper. From (\ref{SpecEqMix2}) we have
\beq
\Big( k_0 + \frac{i}{2}\partial_t
    - \vec \alpha \cdot (\vec k - \frac{i}{2}\vec \nabla)
    - \gamma^0 \hat m_0 - i\gamma^0 \gamma^5 \hat m_5
\Big) \bar G_H = 1 \,,
\label{GHeq2}
\eeq
where again $\bar G_H = G_H \gamma_0$. The only difference between this equation and the corresponding equations for $\bar G^<$ and ${\cal A}\gamma^0$ is the factor of unity on the right hand side.  So we can again make the decomposition 
\beq 
  \bar G_H = \sum_h g_H^h \otimes \frac12(1 + h \hat k\cdot
  \vec \sigma) \,, 
\label{specGH1} 
\eeq
rewrite (\ref{GHeq2}) in component form and separate the equations to constraints and kinetic equations. The only difference in these equations for $g_H^h$ with respect to those for $g_h^<$ occurs in the first constraint equation, which now becomes
\beq
k_0 g^h_{H0} + h|\vec{k}| g^h_{H3} - m_R g^h_{H1} + m_I g^h_{H2} = 1 \,.
\label{CEforgH0}
\eeq
All the other component equations are identical to those for  $g^<_h$ in (\ref{Hset1}). It is straightforward to solve $g_{Hi}^h$ in terms of $g_{H0}^h$, and putting these back to (\ref{CEforgH0}) gives 
\beq
(k_0^2 - \vec{k}^2 - |m|^2) g^s_{H0} = 1 \,.
\label{PPEqn}
\eeq
Now, if we were careful in keeping the $\epsilon$-factors through our computation, we would have found that the function $g^h_{H0}$ solving the equation (\ref{PPEqn}) is in fact the {\em principal part} distribution
\beq
g^h_{H0} = PP \, \frac{1}{k^2 - |m|^2} \,.
\eeq
It is again straightforward to show that the corresponding $4\times 4$-function is given by
\beq
G_H = (\kdag +  m_R - i\gamma^5 m_I) PP \, \frac{1}{k^2 - |m|^2} \,.
\label{Gheq}
\eeq
This form saturates the (momentum space equivalent of the) spectral relation  (\ref{spectrel}) between $G_H$ and ${\cal A}$ and hence is the complete  solution for $G_H$ in the noninteracting mean field limit. The solutions (\ref{specA3}) and (\ref{Gheq}) guarantee that also the retarded and advanced propagators $G^{r,a}$ do not contain any contribution from the coherence solutions.

\subsection{Equilibrium limit for $G^{<,>}$}

The coherence solutions are also excluded from the dynamical functions in the thermal limit. First note that for free fields the equation of motion, and hence the solution, for $G^>$ is identical in form to that for $G^<$. However, the \apriori independent distributions $f^{h\, <,>}_{s_{k_0}}$ in functions $G^<$ and $G^>$ are constrained by the matrix relation (\ref{Hdecomposition}): $G^> + G^< = - 2i {\cal A}$. (Note that we drop the $<,>$-indices on $f^h_\alpha$-functions everywhere where there is no danger of confusion.) This implies that
\beq
f^{h\,<}_{s_{k_0}} + f^{h\,>}_{s_{k_0}} = 1 
\qquad {\rm and} \qquad
f^{h\,<}_{1,2} + f^{h\,>}_{1,2} = 0 \,.
\eeq
These relations hold generically as long as the spectral solutions are valid. Furthermore, in the thermal equilibrium limit functions $G^<$ and $G^>$ are related by the  Kubo-Martin-Schwinger (KMS) boundary condition~\cite{LeBellac}\footnote{Our sign-convention for $G^<$ is opposite to the usual one, see Eqs.~(\ref{GFs},\ref{Gmatrix}), hence the KMS-condition does not involve an explicit - sign.}: 
\beq
G^>_{\rm eq}(t) \equiv G^<_{\rm eq}(t + i\beta)  \qquad \Rightarrow \qquad 
G^>_{\rm eq}(k_0) = e^{\beta k_0}G^<_{\rm eq}(k_0) \,.
\eeq
This matrix condition is strong enough to impose the vanishing of the coherence functions,
\beq
f^{h\,<,>}_{1,2} = 0 
\label{eqf12}
\eeq
and setting the mass-shell distributions to
\beqa
f^{h\,<}_{s_{k_0}} &=& n_{\rm eq}(k_0) \nonumber \\
f^{h\,>}_{s_{k_0}} &=& 1- n_{\rm eq}(k_0)  \,,
\label{eqfspm}
\eeqa
where $n_{\rm eq}(k_0) = 1/(e^{\beta k_0} +1)$ is the usual Fermi-Dirac distribution. Using solutions (\ref{eqf12}-\ref{eqfspm}) in Eq.~(\ref{fullchiralHOMOG}), and summing over the helicities in the decomposition given by Eq.~(\ref{connectionHOMOG}) one finds that:
\begin{eqnarray}
 iG^<_{\rm eq} &=& 
       2\pi {\rm sgn}(k_0) (\kdag +  m_R - i\gamma^5 m_I) \,n_{\rm eq}(k_0) \,
            \delta(k^2-|m|^2) 
\nonumber \\
 iG^>_{\rm eq} &=&  2\pi {\rm sgn}(k_0) (\kdag +  m_R - i\gamma^5 m_I)
         \,  (1 - n_{\rm eq}(k_0)) \, \delta(k^2-|m|^2) \,,
\label{gthermal}
\end{eqnarray}
which are recognized as the standard thermal equilibrium propagators~\cite{LeBellac}.  Generally, in our treatment, the functions $f^{h\,<}_\alpha(|\vec{k}|,t)$ are time-dependent and carry information of both quantum coherence and of statistical out-of-equilibrium conditions.

\section{Weighted density matrix}
\label{sec:weighted}

The dynamical evolution of a free system should be described by the kinetic (AH)-equation in (\ref{Heq}), but the singular structure of $g^<_h(k_0,|\vec{k}|;t)$ complicates the matters. Since distributions are only well defined inside an integral, the equation for $g^<_h$ must be integrated one way or the other to get sensible evolution equations for the on-shell functions $f_{\pm,1,2}$, which are the objects that must carry the physical information about the system. The necessity of such an integration is actually something to be expected, because we can never have a complete information about the variables that define a system under consideration. To quantify this thinking, we introduce the physical density matrix as a weighted average of the original distribution matrix $g^<_h$~\cite{HKR1}:
\beq
  \rho_{\cal W} ( k_0,  |\vec{k}|,  h; t)
     \equiv \sum_{h'} \int \frac{{\rm d}k'_0}{2\pi} \frac{{\rm d}^3
       k'}{(2\pi)^3}\; {\cal W} (k_0, |\vec{k}|, h \mid\hskip-0.5truemm\mid
     k'_0,|\vec{k'}|,h'\,;\,t)
      \; g^<_{h'}(k'_0,|\vec{k'}|;t) \,,
\label{rhoW}
\eeq
where the weight function ${\cal W} (k_0, |\vec{k}|, h \mid\hskip-0.5truemm\mid k'_0,|\vec{k}|',h'\,;\,t)$ encodes our knowledge about the energy, momentum and helicity variables of the state. Our task is now to find out the equations of motion for this weighted density matrix, and consequently for the on-shell functions $f_{\pm,1,2}$. For example, a complete information of the energy the momentum and the helicity of the state immediately renders the evolution of appropriate density matrix trivial one: $\partial_t \rho= 0$, whose solutions are just the constant freely propagating helicity eigenstates~\cite{HKR1}. 

For a more interesting example, where the quantum coherence effects become important, consider a case where we have a precise information of the helicity and size of momentum, but know nothing about the energy. Effectively this means that we cannot differentiate between particles and antiparticles with the same helicity and energy. This is the appropriate situation for computing the particle production during preheating for example, to be discussed in detail in the section \ref{sec:preheating} below. The weight function describing this situation is just
\beq
{\cal W}_1 = \frac{(2\pi)^3}{4\pi \vec{k'}^2} \delta(|\vec{k}|-|\vec{k'}|) \, \delta_{h,h'} \,.
\label{weight2}
\eeq
In this case the weighted density matrix $\rho_{{\cal W}_1}$:
\beqa
\rho_{{\cal W}_1}(k_0,|\vec{k}|,h;t) 
&=& \sum_{h'} \int \frac{{\rm d}k'_0}{2\pi} \frac{{\rm d}^3k'}{(2\pi)^3}\;
\frac{(2\pi)^3}{4\pi \vec{k'}^2} \delta(|\vec{k}|-|\vec{k'}|) \, \delta_{h,h'}
 \; g^<_{h'}(k'_0,|\vec{k'}|;t) 
\nonumber\\ 
&=& \int \frac{{\rm d}k_0}{2\pi} \; g^<_{h}(k_0,|\vec{k}|;t)
\nonumber\\[2mm] 
&\equiv& \; \rho_h(|\vec{k}|;t) \equiv \langle g_h^< \rangle \,,
\label{smearedrhoex}
\eeqa
obeys the evolution equation of same form as $g^<_h$:
\beq
\partial_t \rho_h = -i[H, \rho_h] \,.
\label{rho3}
\eeq
Here the commutator $[H, \rho_h]$ does not vanish, giving rise to a nontrivial time dependence for the nonsingular weighted density matrix $\rho_h$. In particular, the $k_0 = 0$-shell functions $f_{1,2}$ are now directly related to the components of the density matrix and affect the evolution of the mass-shell functions $f_{\pm}$ as well. Using the Bloch-representation $\rho_h \equiv \frac{1}{2}(\langle g^h_0 \rangle + \langle
\vec{g}^h \rangle \cdot \vec{\sigma})$, we have the following relations between $\langle g^h_0 \rangle$ and $f_{\pm,1,2}$:
\begin{eqnarray}
\langle g^h_0 \rangle &=& f^h_+ + f^h_-
\nonumber \\[3mm]
\langle g^h_1 \rangle &=& \frac{m_R}{\omega}(f^h_+ - f^h_-) + f^h_1
\nonumber \\[1mm]
\langle g^h_2 \rangle &=& -\frac{m_I}{\omega}(f^h_+ - f^h_-) + f^h_2
\nonumber \\
\langle g^h_3 \rangle &=& -h\frac{|\vec{k}|}{\omega}( f^h_+ - f^h_-) + h \Big(
\frac{m_R}{|\vec{k}|}f^h_1 - \frac{m_I}{|\vec{k}|}f^h_2\Big) \,,
\label{rhocompbloch}
\end{eqnarray}
where $\omega \equiv \sqrt{\vec{k}^2 + |m|^2}$. 
Note that the with the weight (\ref{weight2}), the physical density matrix $\rho_h$ is just the zeroth moment of the original distribution matrix $g_h^<$ with respect to the energy.
 
\section{Physical quantities}
\label{sec:physical}

Equations (\ref{rhocompbloch}) can be inverted to obtain $f^h_{\pm,1,2}$ in terms of moment Bloch functions $\langle g^h_{\alpha} \rangle$. We are especially interested in the expressions for the mass-shell distributions $f_{\pm}$, defined in Eq.~(\ref{SpecSol}), which in our approach are directly related to the particle and antiparticle number densities. Indeed, according to Feynman-Stuckelberg interpretation  the phase-space particle number density is $n \equiv f_+$, while for fermionic antiparticles $\bar n \equiv 1 - f_-$. With these identifications, using the inverse relations of Eq.~(\ref{rhocompbloch}) we find that for a given 3-momentum $\vec{k}$ and helicity $h$ the out-of-equilibrium particle and antiparticle numbers can be written as:
\beqa
n_{\vec{k}h} &=& \frac{1}{2 \omega}\left(-h|\vec{k}| \langle
  g^h_3 \rangle + m_R \langle g^h_1 \rangle - m_I \langle g^h_2
  \rangle \right) + \frac{1}{2} \langle g^h_0 \rangle 
\nonumber \\ 
{\bar n}_{\vec{k}h} &=& \frac{1}{2 \omega}\left(-h|\vec{k}| \langle
  g^h_3 \rangle + m_R \langle g^h_1 \rangle - m_I \langle g^h_2
  \rangle \right) - \frac{1}{2} \langle g^h_0 \rangle + 1 \,.
\label{partnumber}
\eeqa
Setting ${\rm Tr}(\rho^h) = \langle g^h_0 \rangle \equiv 1$\footnote{Physically this constraint corresponds to setting the chemical potential to zero. Indeed, since $\langle g^h_0 \rangle = f^h_+ + f^h_- = n_{\vec{k}h} - {\bar n}_{\vec{k}h} + 1$, we see that setting $\langle g^h_0 \rangle \equiv 1$ reduces to $n_{\vec{k}h} \equiv {\bar n}_{\vec{k}h}$.} these expressions reduce to the ones obtained in ref.~\cite{Prokopec_partnumber}\footnote{Different signs of terms involving $h$ and $m_I$ are due to a different convention in our definition of the Hermitian Wightmann function Eq.~(\ref{herm_gless}).}, where they were derived using the solutions to a Dirac equation and a Bogolyubov transformation to diagonalize the fermionic Hamiltonian. Indeed, this definition of the particle number in terms of the moment functions was one of the main results of the paper \cite{Prokopec_partnumber}. Here the definition of the particle number is trivial, and we introduced the expressions (\ref{partnumber}) merely to show that our definition does agree with the other approach. 
In a future work~\cite{HKRfuture}, we shall show that in the bosonic case our particle number differs from the one obtained in ref.~\cite{Prokopec_partnumber}, but is consistent with the definition of the ref.~\cite{Berges}.

It is also interesting to see what kind of expressions other physical quantities like energy density and pressure will have in terms of the components $\langle g^h_{\alpha} \rangle$ or the on-shell functions $f_{\pm,1,2}$. These quantities are defined as the ensemble averages the energy momentum tensor\footnote{Here we use the symmetric (Belinfante) version of energy-momentum tensor~\cite{energy_momentum}. However, the same results would have been obtained using the canonical tensor $T^{\mu\nu} = \bar\Psi\left[i\gamma^\mu\partial^\nu - g^{\mu\nu}\left(i\gamma^\mu \partial_\mu - m_R - i\gamma^5 m_I\right)\right]\Psi$.}
\beq
\theta^{\mu\nu}=\frac{i}{4}\Big(\bar\Psi\gamma^\mu\partial^\nu\Psi  
- \partial^\nu\bar\Psi\gamma^\mu\Psi\Big) + \mu \leftrightarrow \nu 
\eeq
and so, for example for the energy density we get
\beqa
\langle{\cal H}(x)\rangle = \langle\theta^{00}(x)\rangle 
&=& \int \frac{{\rm d}^4 k}{(2\pi)^4}\,{\rm Tr}\left[\left(\vec\gamma \cdot \vec
    k + m_R + i m_I \gamma^5 \right)iG^<(k,t)\right] 
\nonumber \\[1mm]
&=& \sum_h \int \frac{{\rm d}^3 k}{(2\pi)^3} \left(-h|\vec{k}| \langle
  g^h_3 \rangle + m_R \langle g^h_1 \rangle - m_I \langle g^h_2
  \rangle \right) 
\nonumber \\
&=& \sum_h \int \frac{{\rm d}^3 k}{(2\pi)^3}\,\omega_{\vec
  k}\left(n_{\vec{k}h}+{\bar n}_{\vec{k}h} -1\right)\,.
\label{energy_dens}
\eeqa
According to expectations the result is simply a sum of free particle and antiparticle contributions. The last term in the last row is the sum of corresponding vacuum energies. For the pressure we get instead
\beqa
\langle P(x)\rangle = \langle\theta^{ii}(x)\rangle 
&=& \int \frac{{\rm d}^4 k}{(2\pi)^4}\,{\rm Tr}\left[\gamma^i k^i
  \,iG^<(k,t)\right] 
\nonumber \\[1mm]
&=& \sum_h \int \frac{{\rm d}^3 k}{(2\pi)^3}\,\frac{1}{3}\left(-h|\vec{k}| \langle
  g^h_3 \rangle \right) 
\nonumber \\
&=& \sum_h \int \frac{{\rm d}^3 k}{(2\pi)^3}
\,\frac{1}{3}\left(\frac{{\vec{k}}^2}{\omega}\left(n_{\vec{k}h}+{\bar
      n}_{\vec{k}h}  - 1\right) - m_R f^h_1 + m_I f^h_2 \right) \,. 
\label{pressure}
\eeqa
Now we see that in addition to normal free particle and antiparticle terms there is an explicit contribution from the coherence shell functions $f_{1,2}$, signalling that at the quantum level the pressure is different from the statistical one. In most cases the quantum effect would be unobservable however, since the coherence functions $f_{1,2}$ are typically oscillatory with microscopic time-scales $\Delta t_{\rm osc} \sim 1/\omega$, so that the  classical thermodynamical pressure arises from Eq.~(\ref{pressure}) when it is averaged out over any time-scales exceeding the quantum scale $\Delta t_{\rm osc}$.

\section{Interacting fields}
\label{sec:interactions}

Having set up the density matrix formalism for treating quantum coherence
phenomena in classical backgrounds, we now wish to extend our work to include 
interactions. To this end we must use the full Kadanoff-Baym equations
(\ref{SpecEqMix1}-\ref{DynEqMix}). In their complete generality these equations
couple nonlinearily the three Green's functions $G^<$, ${\cal A}$ and $G_H$. Solving these equations simultaneously would be an overwhelmingly difficult 
problem however, and we shall adopt a series of approximations to extract 
the relevant physics in what becomes the quasiparticle limit. The key idea in our approach is to divide the problem into two parts. We first need to find a reasonable approximation for the phase space of the problem in terms of on-shell distributions as we did earlier in the case of the free fields. Second, we must find the equations of motion for the on-shell distribution functions including interactions. Looking at equations (\ref{SpecEqMix1}-\ref{DynEqMix}) one immediately sees that the couplings between equations due to terms involving the pole function $G_H$ are causing most problems along the way to any Boltzmann-equation type approximations for the problem. If these can be neglected, the equations for $G^<$ and ${\cal A}$ will decouple and one does not need to solve $G_H$ at all. Fortunately, as we shall see, this indeed is a reasonable approximation in  the weak coupling limit.  As a result of this approximation ${\cal A}$ and $G^<$ will continue to have on-shell solutions, which allows us to go through the procedure leading to our density matrix formalism, yet including the effects of decohering interactions.

\subsection{Quasiparticle approximation}

Let us first consider the weak coupling approximation for the pole equations
(\ref{SpecEqMix1}-\ref{SpecEqMix2}). To zeroth order in gradients, but for
arbitrary $\Sigma_H$ and $\Gamma$ we obtain:
\begin{eqnarray}
G_H &=& \frac{1}{1 + (G_0 \Gamma )^2} \,G_0
\nonumber \\
{\cal A} &=& \frac{1}{1 + (G_0 \Gamma )^2} \,G_0 \Gamma \, G_0 \,,
\label{quasi1}
\end{eqnarray}
where $G_0$ now includes the real part of the self-energy $\Sigma_H$:
\beq
G_0^{-1} = \kdag - m_R - im_I\gamma^5 - \Sigma_H \,,
\eeq
and $\Gamma$ is defined in equation (\ref{gamma}). We see that ${\cal A}$ no more has a spectral solution, and the phase space  is truly 4-dimensional. However, ${\cal A}$ clearly reduces to a spectral form in the limit $\Gamma \rightarrow 0$\footnote{Note that $\Gamma$ is in fact a $4\times 4$ matrix operator, so with the limit $\Gamma \rightarrow 0$ we mean that the coupling $y\rightarrow 0$. Actually we have to keep a finite, but arbitrarily small imaginary part in the definitions of propagators, due to the inclination of the Keldysh path in the complex time plane. Thus in the collisionless limit
$\Gamma \rightarrow {\rm sgn}(k_0)\epsilon 1_4$.}:
\begin{eqnarray}
{\cal A} &\rightarrow & \pi\,{\rm sgn}(k_0) \delta(G_0^{-1})
\nonumber \\
         & = & \pi\,{\rm sgn}(k_0) G_0 \, {\rm det}(G_0^{-1}) \, \delta ({\rm det}(G_0^{-1})),
\end{eqnarray}
where the determinant inside the delta-function now gives rise to a modified 
dispersion relation:
\beq
{\rm det}(\kdag - m_R - im_I\gamma^5 - \Sigma_H ) = 0.
\label{disprelqp}
\eeq
For the pole-function $G_H$ we would obviously get the corresponding
principal value as in noninteracting case (see section
\ref{sec:spectral}). The limit of taking $\Gamma\rightarrow 0$, while
keeping $\Sigma_H$ finite is just the well known {\em quasiparticle  approximation}. It is often a reasonable approximation to be made in
the weak coupling limit. Technically  one should require that
$\Sigma_H$ is of lower order in the coupling constants than is
$\Gamma$, and this often indeed is the case: for gauge interactions
for example one finds $\Sigma_H \sim g^2$ and $\Gamma \sim g^4$ in the lowest 
order in the gauge coupling $g$.  However, even when
the coupling hierarchy is not there, the quasiparticle limit can be a
useful first approximation.

\subsection{KB-equation with collisions}
\label{sec:KB_collisions}

We now turn our attention to the dynamical equation (\ref{DynEqMix}), which
in the mean field limit becomes:
\beq
  (\kdag - \frac{i}{2} \deldag_x - m_R
       - i m_I \gamma^5 - \Sigma_H ) \, G^<  -  \Sigma^< G_H
      = {\cal C}_{\rm coll} \,,
\label{DynEqMixMF}
\eeq
where
\beq
{\cal C}_{\rm coll} = \frac{1}{2} (\Sigma^> G^< - \Sigma^< G^>) \,.
\label{collMF}
\eeq
Clearly the $G_H$-mixing term prevents Eq.~(\ref{DynEqMixMF}) from providing unique solution for $G^<$ even when the spectral function is known. However, neglecting this term is in fact consistent with the quasiparticle approximation. One can see this from the fact that Eq.~(\ref{SpecEqMix1}) for the spectral function ${\cal A}$ can be obtained as a sum of the evolution equations for $G^<$ and $G^>$, and that the mixing term $\sim \Gamma G_H$ in this equation arises from the sum of $G_H \Sigma^{<,>}$-terms in the equations for $G^{<,>}$. Since neglecting the $\Gamma$-mixing was precisely what defined the quasiparticle approximation for the pole-equations, corresponding terms should be discarded in the equations for $G^{<,>}$ as well. Thus, in the quasiparticle and mean field limit we have
\beq
  (\kdag - \frac{i}{2} \deldag_x - m_R
       - im_I \gamma^5 - \Sigma_H) \, G^<
      = {\cal C}_{\rm coll} \,.
\label{DynEqMixQPA}
\eeq
The collision term can be written in terms of $\Gamma$ and ${\cal A}$ as 
follows: 
\beq
{\cal C}_{\rm coll} = -i \Gamma G^< + i{\Sigma}^< {\cal A}\,,
\eeq
so that given a solution for ${\cal A}$ equation (\ref{DynEqMixQPA}) can 
be solved in weak coupling limit to give $G^<$. The overall strategy
is the same as before: one  first divides equations into
kinetic equations and constraints.  Consistency with the quasiparticle approximation requires that one must neglect all collision terms in the constraint  equations, since including them would give corrections that are of the same  order as terms neglected in quasiparticle approximation. 
When this procedure is followed the constraint equations for $G^<$ lead to 
the same quasiparticle mass-shell solutions for $G^<$ as for the spectral function, plus the additional solutions describing the coherence at $k_0=0$-shell, because for $G^<$ the latter are not suppressed by the spectral sum-rule. However, our main interest in this paper is to study {\em the effects of decohering collisions} on the quantum coherence of the system. Because these do
not qualitatively depend on the modifications to dispersion relations, we will 
set $\Sigma_H$ to zero for simplicity in what follows. This obviously 
reduces the functions ${\cal A}$ and $G_H$ to their collisionless limits 
given in Eqs.~(\ref{specA3},\ref{Gheq}), and the kinetic equation 
eventually becomes simply
\beq
\Big( k_0 + \frac{i}{2}\partial_t
    - \vec \alpha \cdot (\vec k - \frac{i}{2}\vec \nabla)
    - \gamma^0 \hat m_0 - i\gamma^0 \gamma^5 \hat m_5
\Big) \bar G^<
      = i\gamma^0{\cal C}_{\rm coll}\gamma^0 \,.
\label{DynEqMixQPAFinal}
\eeq
Until now our analysis has been entirely independent of the particular type of interactions. However, it is not possible to analyse the problem further before explicitly defining the interaction terms. Before doing that we shall notice that whatever the form of the interaction, if the self-energies $\Sigma^{<,>}$ are thermal obeying the Kubo-Martin-Schwinger (KMS) relation $\Sigma^>(k,x) = e^{\beta k_0} \Sigma^<(k,x)$, the collision term reduces to:
\beq
{\cal C}_{\rm coll} 
= -i\,\Gamma_{\rm KMS} \left(G^< - G_{\rm eq}^<\right)\,.
\label{collision_thermal}
\eeq
This follows from the KMS-relation combined with the relation $i G_{\rm eq}^< = 2\,n_{\rm eq}(k_0){\cal A}$ given by Eqs.~(\ref{specA3}) and (\ref{gthermal}). The form of collision term (\ref{collision_thermal}) is familiar from relaxation-time approximation and it yields the relaxation to thermal equilibrium in time scale $1/\Gamma$ in the absence of any driving terms.

After the interactions are specified and an explicit form of $\Gamma$ is known we can proceed to write down our equations of motion. These equations will be generalizations of the equations (\ref{Heq}) in the free field case.  As we mentioned above, the constraint equation (H) will be untouched because we are working in the quasiparticle mean field limit and we neglect the term $\Sigma_H$ for simplicity.  The kinetic equation (AH) will receive contributions from the collision terms however. It is very important to realize, and this can already be seen from equation (\ref{DynEqMixQPAFinal}) that as the singular solution for the function $G^<$ is introduced into the kinetic equations, we will encounter projections of the collision term $\Gamma$ on all different shells contributing to $G^<$. That is, the external momentum configuration entering to the evaluation of $\Gamma$ will depend on the particular shell multiplying it in the collision integral.  In particular, the coherence shells will pick interaction terms with $\Gamma(k_0=0)$, which in general are completely different from the terms involving the usual mass-shell functions.

\subsection{Computation of the self-energies}
\label{sect:self-energies}

\begin{figure}
\centering
\includegraphics[width=0.5\textwidth]{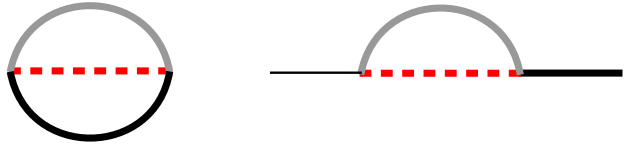}
    \caption{Diagrams contributing to the 2PI-action and the self-energy 
              at the one loop level due to interaction (\ref{interaction}).}
    \label{fig:self1a}
\end{figure}

In this subsection we give an explicit evaluation of the self-energy 
corrections arising from a specific form of the interaction. This is a 
rather technical calculation whose results are particular to the chosen
interaction. A reader not interested in these details can skip this subsection 
and move directly to the next section where we will continue developing the 
kinetic equations given the results from this subsection.  We choose to consider 
the following left-chiral non-diagonal Yukawa interaction term:
\beq 
{\cal L}_{\rm int} = - y\; \bar \psi_L \phi \, q_R + h.c. 
\label{interaction}
\eeq
where $\psi$ is the considered fermion (quark) field, $q$ is some other fermion field and $\phi$ is a complex scalar field. As mentioned before we use the two-particle irreducible (2PI) action method to calculate the self-energies (\ref{2PIsigmas}). The lowest order 2PI-graph based on interaction (\ref{interaction}) is presented in Fig.~\ref{fig:self1a}a, and it gives the contribution 
\beq 
\Gamma_{\rm 2PI} = -|y|^2 \int_C d^4u\,d^4v\, {\rm Tr}\left[P_R G_q(u,v) P_L
                           G(v,u)\right]\Delta(u,v) \,, 
\label{gamma2pI}
\eeq
where $G$, $G_q$ and $\Delta$ are propagators of the considered fermion, 
quark and scalar field and the integration is over the Keldysh path. From this
we get fermion $\psi$ self-energies:
\beq
   \Sigma^{ab}(u,v) =  -iab \frac{\delta \Gamma_2[G]}{\delta
  G^{ba}(v,u)} = i |y|^2 P_R G_q^{ab}(u,v) P_L \Delta^{ab}(u,v)
\label{sigma-ab2}
\eeq   
and in particular\footnote{The 2PI-formalism is not necessary for obtaining
these results. Equivalently with Eq.~(\ref{gamma2pI}) one can directly write 
down the self-energy with complex time variables:
\beq
\Sigma_{\cal C}(u,v) = i|y|^2 P_R G_{\cal C}^q(u,v) P_L 
                                \Delta_{\cal C}(u,v) \,,
\eeq
from which (\ref{sigma-ab2}) readily follows by making the appropriate
choices for the position of variables on the time path.
}
\beq 
  \Sigma^{<,>}(u,v) = i |y|^2 P_R G_q^{<,>}(u,v) P_L \Delta^{<,>}(u,v)\,.
\eeq
We obviously need these quantities in the mixed representation and 
after a Wigner transformation the self-energies become:
\beq \Sigma^{<,>}(k,x) = i |y|^2 \int
\frac{d^4k'}{(2\pi)^4} P_R G_q^{<,>}(k',x) P_L \Delta^{<,>}(k-k',x) \,. 
\label{self1}
\eeq 
To keep things simple, we will now assume that the quark- and scalar 
distributions appearing in the loop are thermal. The appropriate quark
propagators (with real constant mass) can then be read directly from 
(\ref{gthermal}):
\beqa
  iG_{q,{\rm eq}}^<(k) &=& 
   2\pi \left( \kdag + m_q \right) 
        \delta(k^2-m_q^2) {\rm sgn}(k_0) n_{\rm eq}(k_0)         
\nonumber \\ 
 iG_{q,{\rm eq}}^>(k) &=& 
   2\pi \left( \kdag + m_q \right) 
         \delta(k^2-m_q^2) {\rm sgn}(k_0) (1 - n_{\rm eq}(k_0)) \,,
\label{thermal_prop_quark}
\eeqa
while the equivalent expressions for the bosonic propagators are~\cite{PSW}
\beqa
i\Delta_{\rm eq}^<(k) &=& 
    2\pi \delta(k^2-m_\phi^2) {\rm sgn}(k_0) n^\phi_{\rm eq}(k_0) 
\nonumber \\ 
i\Delta_{\rm eq}^>(k) &=& 
    2\pi \delta(k^2-m_\phi^2) {\rm sgn}(k_0) (1 + n^\phi_{\rm eq}(k_0)) \,,
\label{thermal_prop_delta}
\eeqa
where the thermal fermion and boson distribution functions $n_{\rm eq}$ and $n^\phi_{\rm eq}$ are 
\beq 
 n_{\rm eq}(k_0) = \frac{1}{e^{\beta k_0} + 1} \,, \qquad 
 n^\phi_{\rm eq}(k_0) = \frac{1}{e^{\beta k_0} - 1} \,. 
\eeq
In the thermal approximation the explicit $x$-dependence of the propagators
$G_q^{<,>}$ and $\Delta^{<,>}$ drops, and it might appear that also
$\Sigma^{<,>}$ then becomes $x$-independent. That this is not so follows from
the nontrivial dependence of $\Delta^{<,>}_{\rm eq}(k-k')$ on the exteral 
4-momentum $k$. When $k$ enters to the on-shell delta-functions it introduces 
dependence on the mass of the external field $\psi$. Computing $\Sigma^{<,>}$
is now a simple matter of substituting thermal propagators
(\ref{thermal_prop_quark}-\ref{thermal_prop_delta}) to the expression for 
the self energy (\ref{self1}) and using the delta-functions to perform as 
many of the integrals as possible.  Before doing the actual computation let
us note however, that in the thermal limit $\Sigma^{<,>}$ can only depend on
two independent 4-vectors: $k^\mu$ and the 4-velocity of the plasma
$u^\mu$. Given the chiral structure of the interaction the most general form 
for any $\Sigma^{ab}$ is thus
\beq
\Sigma^{ab}(k) = \big( \, A^{ab} \kdag + B^{ab} \udag \, \big) \; P_L \,,
\label{chiralform}
\eeq
where the isotropy of the thermal distribution implies that functions $A^{ab}$ and $B^{ab}$ can only depend on $k_0$, $|\vec k|$. Note that at this point $\Sigma^{ab}$ is actually independent of the external coordinate $x$; the $x$-dependence (actually only $t$-dependence in this paper) is only introduced through projections to mass shells where the dispersion relation $k_0 = \pm\omega_k(x)$ depends on $x$ through the mass of the $\psi$-field.  Note also that Eq.~(\ref{chiralform}) is more general than the specific interaction we 
are studying here: it is valid for any $L$-chiral interaction and to any order 
in the loop expansion as long as the loop particles are assumed to be thermal. 
Substituting the expressions (\ref{thermal_prop_quark}-\ref{thermal_prop_delta}) to Eq.~(\ref{self1}) we get, after integration over $k_0'$:
\beqa 
  i\Sigma^<(k) &=& 
      2\pi |y|^2 \sum_\pm \int 
        \frac{d^3k'}{(2\pi)^3 4 \omega^q_{k'}\omega^\phi_{k-k'}} 
         n_{\rm eq}(\pm \omega^q_{k'}) n^\phi_{\rm eq}(k_0 \mp \omega^q_{k'}) 
         \Slashnew{k}'_\pm P_L 
\nonumber \\ &&
\hspace{20mm} \times\left(\delta(k_0 \mp \omega^q_{k'} -
   \omega^\phi_{k-k'}) - \delta(k_0 \mp
   \omega^q_{k'} + \omega^\phi_{k-k'}) \right), 
\label{self2}
\eeqa
where $\omega^{q,\phi}_{p} \equiv \sqrt{\vec{p}^2 + m_{q,\phi}^2}$
and $k'_\pm  \equiv (\omega^q_{k'}, \pm \vec{k'})$. The angular integrals 
in the expression (\ref{self2}) can be further evaluated in spherical
coordinates with the result:
\beq 
  \Sigma^<(k) = 
         \left[\Sigma^<_0 \,\gamma^0 - \Sigma^<_3 \,
                      \big(\hat{k}\cdot \vec{\gamma}\big) \right] \, P_L \,, 
\label{self3}
\eeq
where $\hat k = \vec k/|\vec{k}|$ and
$i\Sigma^<_{0,3} = i\Sigma^<_{0,3}(k_0,|\vec{k}|)$ are real-valued
functions of the phase space. For our present analysis we need to evaluate the
self-energies $\Sigma^{<,>}$ both on the mass-shell $k_0^2-\vec{k}^2=|m(x)|^2$
as well as on the $k_0=0$-shell. On the mass-shell we get:
\beqa 
i\Sigma^<_0(k_0=\pm\omega_k(x),|\vec{k}|) 
   &=& \frac{|y|^2 T^2}{8\pi |\vec{k}|} |I_1(k_0,|\vec{k}|)| \,, 
\\ 
i\Sigma^<_3(k_0=\pm\omega_k(x),|\vec{k}|) 
   &=& \frac{|y|^2 T^2}{8 \pi |\vec{k}|}
          \left[\frac{k_0}{|\vec{k}|} \left(|I_1(k_0,|\vec{k}|)| 
               - \frac{|\alpha| \,|m|^2}{k_0^2} |I_0(k_0,|\vec{k}|)|\right)
          \right] \,, 
\label{Sigma_0,3}
\eeqa    
with
\beq 
I_n(k_0,|\vec{k}|) 
  = \theta(\lambda) \int_{\alpha - \beta}^{\alpha + \beta} {\rm d}y \: y^n 
       \frac{1}{(e^y + 1) (e^{k_0/T - y} - 1 )} \,.
\label{I_n}
\eeq  
and
\beqa
\alpha &=& \frac{|m|^2+m_q^2-m_\phi^2}{2|m|^2} \frac{k_0}{T}
       = \frac{\sqrt{s}E_*k_0}{|m|^2 T}
\nonumber \\
\beta  &=& \frac{\lambda^{1/2}(|m|^2,m_q^2,m_\phi^2)}{2|m|^2} 
               \frac{|\vec{k}|}{T}
       = \frac{\sqrt{s}p_*|\vec{k}|}{|m|^2 T} \,,
\eeqa
where $\omega_k \equiv \sqrt{\vec{k}^2+|m|^2}$ and $E_*$ and $p_*$ are
the energy and momentum of the decay products in the decay frame, $\sqrt{s}$ is the invariant mass of the decaying (heaviest) particle and $\lambda(a,b,c) = (a+b-c)^2-4bc$ is the usual kinematic phase space function. The peculiar appearance of the Heavyside step-function with the argument $\lambda \equiv \lambda(|m|^2,m_q^2,m_\phi^2)$  causes these expressions to automatically take care of the correct mass  hierarchy of the fields; the mass $m=m(x)$ may be varying in spacetime, so $\psi$ can for example change from being the lightest to the heaviest of the three fields. On the $k_0=0$-shell we get instead:
\beqa 
i\Sigma^<_0(k_0=0,|\vec{k}|) 
   &=& \frac{|y|^2 T^2}{8\pi |\vec{k}|}
   \int_{\frac{\lambda^{1/2}}{2|\vec{k}|T}}^{\infty} {\rm
     d}y\,\frac{y}{\sinh(y)} 
\\ 
i\Sigma^<_3(k_0=0,|\vec{k}|) 
   &=& 0 \,, 
\label{Sigma_0,3_k0}
\eeqa   
Now $\lambda \geq 0$ always so there is no need for an explicit
step-function. One crucial difference between the mass-shell and the 
$k_0=0$-shell self energies is that the latter are completely
$x$-independent while the former may be $x$-dependent, because
of the possibly $x$-dependent mass, as mentioned earlier.

Because we are computing $\Sigma^{ab}$'s in the thermal limit, the expression 
for $\Sigma^>$ can be obtained from that for $\Sigma^<$ by use of the 
Kubo-Martin-Schwinger (KMS) relation:
\beq 
\Sigma^>(k) = e^{\beta k_0} \Sigma^<(k) \,. 
\label{KMSrelaatio}
\eeq 
This relation can be seen to emerge from (\ref{self2}), where we should obtain $\Sigma^< \rightarrow \Sigma^>$ by replacing $n_{\rm eq}^q \rightarrow 1 - n_{\rm eq}^q$ and $n^\phi_{\rm eq} \rightarrow 1+ n^\phi_{\rm eq}$. KMS-relation 
(\ref{KMSrelaatio}) follows then immediately by use of the relations $1-n_{\rm eq}(k_0) = e^{\beta k_0}n_{\rm eq}(k_0)$ and $1+n^\phi_{\rm eq}(k_0) = e^{\beta k_0}n^\phi_{\rm eq}(k_0)$. Self-energy components  $\Sigma_{0,3}^i$ of course obey the KMS-relation separately. Moreover, we find that
\beq
\Gamma(k) = \frac{i}{2}(1+e^{\beta k_0} ) \Sigma^<(k) \,.
\label{gammakms}
\eeq
Expressions (\ref{self3}) and (\ref{gammakms}) complete the computation of all required self-energy functions needed for the further analysis of Eq.~(\ref{DynEqMixQPAFinal}).

\subsection{Evolution equation for the density matrix}

Using the particular form of the self energy (\ref{self3}) and the KMS-relation
(\ref{gammakms}), which yield a thermal collision term of the type
Eq.~(\ref{collision_thermal}), and the helicity block-diagonal decomposition (\ref{connectionHOMOG}) for $\bar G^<$, the equation (\ref{DynEqMixQPAFinal})
reduces to the following matrix equation for the chiral part $g_h^<$:
\beq
\Big( k_0 + \frac{i}{2}\partial_t - H \Big) g_h^< 
 = -i D \left(g^<_h - (g^<_h)_{\rm eq}\right)\,,
\label{DynEq_collHOMOG}
\eeq
where $H = - h |\vec{k}| \rho^3 + m_R \rho^1 - m_I \rho^2$, as earlier,
\beq
 D \equiv \frac{1}{2}(1+\rho_3) \,\Gamma_h 
 \qquad {\rm with} \qquad \Gamma_h \equiv \Gamma_0 - h\Gamma_3  
\label{D_def_eqn}
\eeq
and $(g^<_h)_{\rm eq}$ is the thermal equilibrium limit of $g^<_h$ defined in Eq.~(\ref{fullchiralHOMOG}). Taking the Hermitian and anti-Hermitian parts of equation (\ref{DynEq_collHOMOG}), and neglecting collisions in the Hermitian equation, consistently with the quasiparticle approximation, we find a generalization of equations (\ref{Heq}) to the case with collisions:
\beqa
{\rm (H):} && \quad 2k_0 g^<_h = \{H, g^<_h\} 
\nonumber \\
{\rm (AH):} &&\quad \partial_t g^<_h = -i[H, g^<_h] - \left\{D , \, g^<_h - (g^<_h)_{\rm
    eq} \right\} \,.
\label{rho_collHOMOG2}
\eeqa
The only difference to the free field case is the appearance of an anticommutator in the anti-Hermitian equation containing the interaction matrix $D$. The precise form of the matrix $D$ in chiral indices will depend on the form of the interactions, but the generic form of the equations in (\ref{rho_collHOMOG2}) including an anticommutator with a generic matrix $D$ is universal to the quasiparticle mean field limit when the self-energy is computed in thermal approximation. The simple form of $D$ in Eq.~(\ref{D_def_eqn}) reflects the chirality of the interaction: $(1+\rho_3)/2$ is just the 2-dimensional version of the left-chirality projector.

Just as in the noninteracting case, the (AH)-equation is ill-defined without an integration procedure. Here we are interested in the evolution of the weighted density matrix relevant for the particle production during preheating in the homogenous time-dependent background.  The momentum and helicity remain to be good quantum numbers even in the presence of the interactions, and so we can take our physical density matrix to be diagonal in these variables. Furthermore, since we have no \apriori information of the energy of the relevant solutions we impose a flat weight on the energy variable. That is, we integrate the (AH)-equation in (\ref{rho_collHOMOG2}) with the weight (\ref{weight2}) introduced in the section \ref{sec:weighted}: ${\cal W}_1 = \frac{(2\pi)^3}{4\pi \vec{k'}^2}\delta(|\vec{k}|-|\vec{k'}|) \, \delta_{h,h'}$. After some manipulations we get the new equation of motion for the physical density matrix $\rho_h$ defined in Eq.~(\ref{smearedrhoex}) including interactions:
\beq
\partial_t \rho_h = -i[H, \rho_h] - I_g \,,
\label{rho_coll_intHOMOG}
\eeq
where
\beqa
I_g &\equiv& \int \frac{{\rm d}k_0}	{2\pi}
      \left\{D \,, \, g^<_h - (g^<_h)^{\rm eq} \right\}
\nonumber \\[3mm]
&\; = \;&
       \Gamma_{m0} \,
              \left( \begin{array}{cc}
                   (f_0 - f_0^{\rm eq}) - h\frac{k}{\omega} (f_3 -
                   f_3^{\rm eq}) &
                   \quad \frac{m}{2\omega}(f_3 -  f_3^{\rm eq}) \\[2mm] 
                   \frac{m^*}{2\omega}(f_3 -  f_3^{\rm eq}) & \quad 0
                      \end{array} \right)
     \nonumber \\[3mm]
          &-& h\Gamma_{m3}
                  \left( \begin{array}{cc}
                   (f_3 - f_3^{\rm eq}) - h\frac{k}{\omega}(f_0 -
                   f_0^{\rm eq}) & \quad \frac{m}{2\omega}(f_0 -
                   f_0^{\rm eq}) \\[2mm] \frac{m^*}{2\omega}(f_0 -
                   f_0^{\rm eq}) & \quad 0
                         \end{array} \right) 
     \nonumber \\[3mm]
          &&+ \,\, \Gamma_{00} 
                  \left( \begin{array}{cc}
                   h\frac{m_R}{k}f_1 -
                   h\frac{m_I}{k}f_2 & \quad
                   \frac{1}{2}(f_1 - if_2) \\[2mm]
                   \frac{1}{2}(f_1 + if_2) & \quad 0
                         \end{array} \right) 
             \,,
\label{coll_int}
\eeqa

\hskip 0.5truecm

\noindent
where we use shorthand notations $f_0 \equiv f_+ + f_-$ and $f_3 \equiv f_+ - f_-$, and similarly for $f_{0,3}^{\rm eq}$ with $f_\pm^{\rm eq} \equiv n_{\rm eq}(\pm \omega)$. Yet we have denoted $k \equiv |\vec{k}|$, and we have dropped the helicity index $h$ in the superscripts of all $f_\alpha$ for convenience. The $\Gamma$-functions appearing as coefficients are now having contributions from different shells on the phase space as we predicted at the end of the section \ref{sec:KB_collisions}. We have used the definitions:
\beqa
\Gamma_{m(0,3)}(|\vec{k}|,t) &\equiv& \Gamma_{0,3}(k_0 = \omega(t),|\vec{k}|) \qquad\quad
(\textrm{positive mass-shell}) \nonumber \\
\Gamma_{00}(|\vec{k}|) &\equiv& \Gamma_0(k_0 = 0, |\vec{k}|) \qquad\qquad \;\;\
(\textrm{$k_0 = 0$ -shell})\,.
\label{gamma_shell}
\eeqa
These are the only independent functions, since $\Gamma$ on the negative mass-shell is related to that on the positive with $\Gamma_{0,3}(-k_0,|\vec{k}|) = \pm \Gamma_{0,3}(k_0,|\vec{k}|)$. Further, $\Gamma_3$ vanishes on the $k_0 = 0$ -shell. It is because of the different values of $\Gamma$-functions in different phase space shells, that the integrated equation of motion (\ref{rho_coll_intHOMOG}) cannot be written in an equally simple matrix form as the original one (\ref{rho_collHOMOG2}).  We have written the collision integral (\ref{coll_int}) in terms of the on-shell functions $f_{\alpha}$ to get the simplest possible expression. Of course, in order to perform any practical calculations one has to use the relations between $\rho_{ij}$ and $f_\alpha$, which can be found using the relations (\ref{rhocompbloch}). In fact, in our numerical calculations in sections \ref{sec:preheating} and \ref{sec:decoherence} we found it easiest to express all quantities in terms of the Bloch-components $\langle g^h_\alpha \rangle$.  Again, the form (\ref{rho_coll_intHOMOG}) is generic for a density matrix defined by the weight (\ref{weight2}), but the precise form $I_g$ strongly depends on the interaction; the more complicated the interaction, the messier $I_g$ becomes. However, there is always a well defined matrix expression for $I_g$ which in the thermal limit is linear in the on-shell functions, involving projections of the interaction matrix components on the positive mass shell and the $k_0=0$-shell.

The generic features of the thermalization and decoherence effects due to collisions are a bit obscure in Eq.~(\ref{rho_coll_intHOMOG}) because of the complex matrix stucture of the collision integral (\ref{coll_int}). They become much simpler if one looks directly at the evolution equations of the on-shell functions $f_\alpha$ in the constant mass limit. Using the relations (\ref{rhocompbloch}) to express $\rho_{ij}$ in terms of $f_\alpha$, we find that in this limit: 
\beqa
\partial_t f_\pm &=&  -\Big(1 -
  h\frac{k}{\omega}\Big)\Gamma_{m0}(f_\pm - f_\pm^{\rm eq}) +
\ldots \nonumber \\[1mm]
\partial_t f_1 &=& -2h\left[-k f_2 + m_I(\frac{m_R}{k}f_1 -
  \frac{m_I}{k}f_2)\right] - \Gamma_{00}f_1 + \ldots \nonumber \\[1mm]
\partial_t f_2 &=& -2h\left[k f_1 + m_R(\frac{m_R}{k}f_1 -
  \frac{m_I}{k}f_2)\right] - \Gamma_{00}f_2 + \ldots \,,
\label{comp_eq_constantHOMOG}
\eeqa
where we have written down only the diagonal interaction terms. Since
all of those are negative ($\Gamma_{m0}, \Gamma_{00} \geq 0$), 
we can immediately see the tendency of interactions to thermalize the mass-shell functions by setting $f_\pm \rightarrow  f_\pm^{\rm  eq}$ as well as to give rise to decoherence $f_{1,2} \rightarrow 0$. The latter effect in particular gives a {\em smooth} damping of the quantum coherence as a result of collisions over a characteristic time scale $1/\Gamma_{00}$. 

\section{Applications}

Let us next study the effects of collisions numerically in two different physical examples. First we consider particle production at the preheating of inflation including the effects of collisions on the coherent particle number  creation. Second, we study the approach to the thermal equilibrium of a quantum system including both thermalization and decoherence effects due to collisions. In both of these examples the 3-momentum $\vec{k}$ and the helicity $h$ are good quantum numbers so that the weight function ${\cal W}_1$ defined in Eq.~(\ref{weight2}), and the resulting equations (\ref{rho_coll_intHOMOG}-\ref{gamma_shell}) with relations~(\ref{rhocompbloch}) are the appropriate one to use in these calculations.

\begin{figure}
\centering
\includegraphics[angle=270,width=0.9\textwidth]{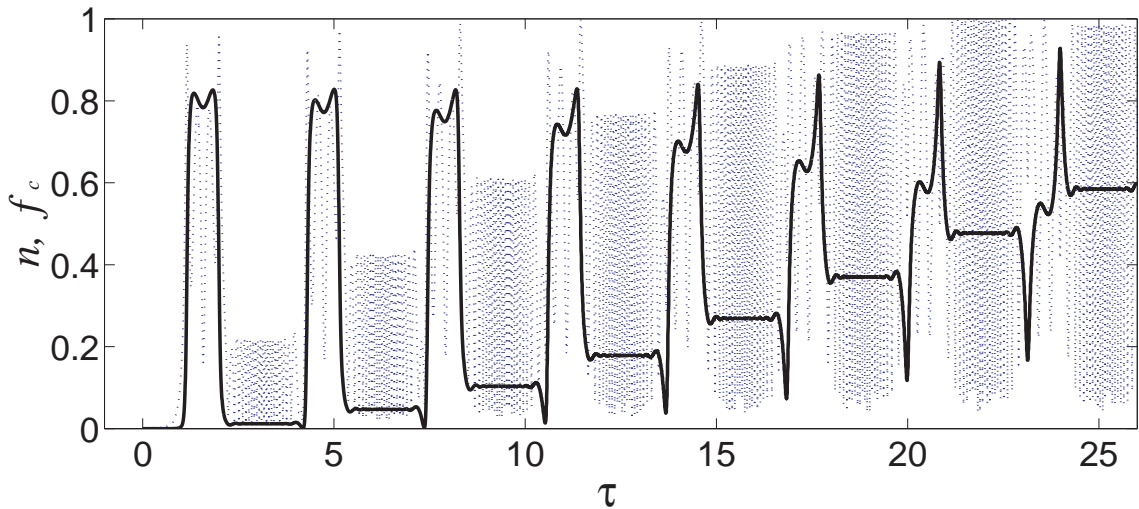}
\caption{Shown is the mean field number density $n_{{\vec k} h}$ of produced fermions as a function of $\tau \equiv \omega_{\varphi} t$ for negative helicity state $h = -1$ in free case (thick black line). 
Also shown is (thin dotted blue line) a function $f^h_c\equiv ((f^h_1)^2 + (f^h_2)^2)^{1/2}$, which describes the overall amount of correlation between fermions and antifermions. Effects of inflaton oscillation are modeled by varying mass $ m(t) = (10 + 15 \cos(2 \omega_{\varphi} t) + i \sin(2 \omega_{\varphi} t))| \vec k |$ where $\omega_{\varphi}= |\vec k |$ is the frequency of the inflaton oscillation. At $\tau = 0$ the fermion system is taken to be an uncorrelated vacuum state.}
\label{fig:particle0}
\end{figure}

\subsection{Particle production at preheating}
\label{sec:preheating}
As our first application we consider the particle production at the preheating stage at the end of the inflation, where a time-dependent fermionic mass is generated by an oscillating inflaton condensate. This system is appropriately described by equations  (\ref{rho_coll_intHOMOG}-\ref{gamma_shell}), assuming that interactions are modelled by our left-chiral Lagrangian Eq.~(\ref{interaction}). We found it most convenient to write the equations entirely in terms of the moment Bloch functions $\langle g^h_\alpha \rangle$ in the code, using relations (\ref{rhocompbloch}), and by computing the particle and antiparticle numbers in the end through Eqs.~(\ref{partnumber}). This preheating problem was considered recently in ref.~\cite{Prokopec_partnumber} in the collisionless limit, and to facilitate the comparison, we will adopt their model for the oscillation of the inflaton condensate. A simple cosine function for inflaton gives rise to the following fermionic mass function: 
\beq
m(t) = m_0 + A \cos(2\omega_\varphi t) + i B\sin(2 \omega_\varphi t),
\eeq
where $m_0$, $A$, $B$ and $\omega_\varphi$ (the inflaton oscillation frequency) are real constants. Let us first consider the noninteracting case with $\Gamma=0$. We assume that there is no initial chemical potential so that $n_{\vec{k}h} \equiv {\bar n}_{\vec{k}h}$ and $\langle g^h_0 \rangle \equiv 1$. This initial condition is preserved throughout the free field evolution, since by Eq.~(\ref{rho3}) one can show that $\partial_t\langle g^h_0 \rangle = 0$. The results of our calculations for helicity $h=-1$ are presented in Fig.~\ref{fig:particle0}. We see that the produced particle number increases steadily as a function of time. This increase takes place at the resonance peak areas while between the peaks the particle number is essentially constant. To arrive to this picture with a {\em parametrical resonance} needs some fine-tuning between the parameters of mass oscillation and the size of momentum $|\vec{k}|$ and helicity $h$; indeed for the opposite helicity $h=+1$ with the otherwise same parameters we would get a completely different figure without a clear resonance behavior. This evolution of the particle number agrees with the results in ref.~\cite{Prokopec_partnumber}. We show also the evolution of the coherence by plotting a function $f_c \equiv \sqrt{f_1^2 + f_2^2}$ (dotted blue line) in Fig.~\ref{fig:particle0}. We see that the particle production is accompanied by a steady growth of coherence. Most of the growth takes place at the resonance peaks, whereas outside the peaks the coherence oscillates with a constant amplitude. Already after a few resonance crossings the coherence has saturated to a maximum.
\begin{figure}
\centering
\includegraphics[angle=270,width=0.9\textwidth]{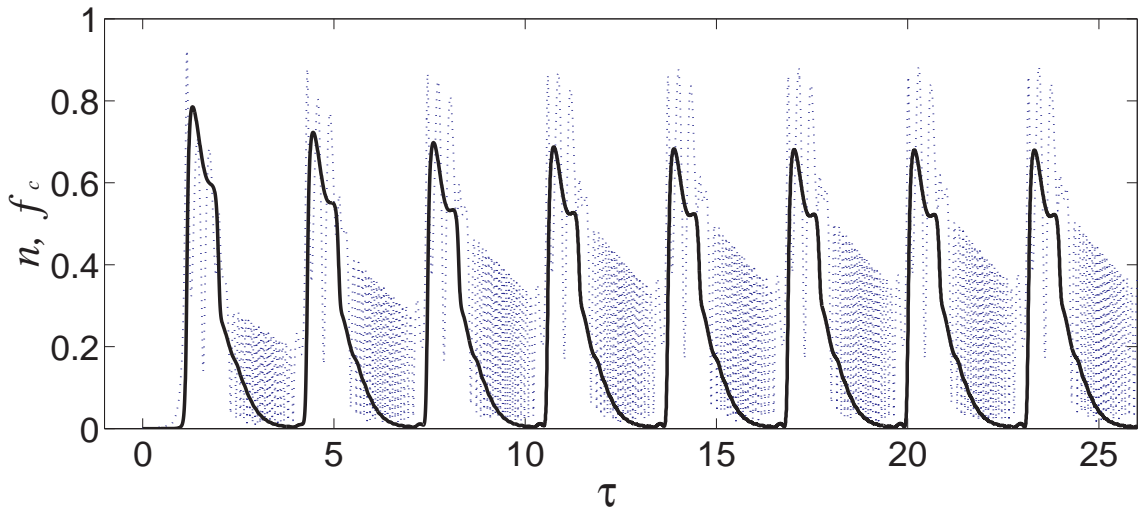}
\caption{The mean field number density $n_{{\vec k} h}$ of fermions as a function of $\tau$ for an interacting field (thick black line), in same setting as in Fig.~\ref{fig:particle0}. Thin dotted blue line again shows the coherence function $f^h_c$ giving the correlation between fermion and antifermion fields. The interaction is taken to be of the form computed in the section \ref{sect:self-energies}, with a thermal background in temperature $T= |\vec k |$ and parameters $y=5, m_q = 0.02|\vec k|$ and $m_\phi =0.1|\vec k|$.}
     \label{fig:particle2}
\end{figure}

\begin{figure}
\centering
\includegraphics[angle=270,width=0.9\textwidth]{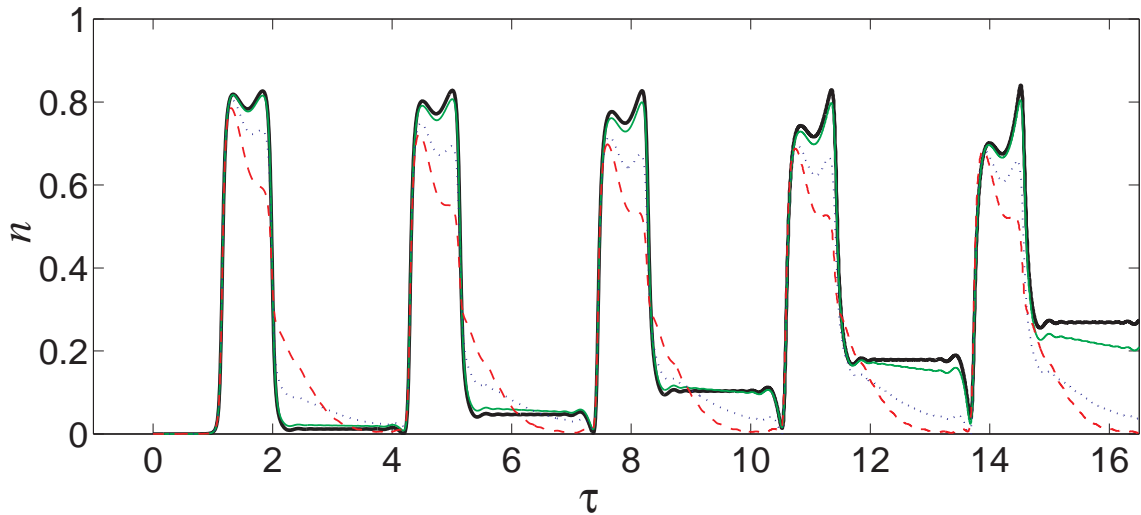}
\caption{Shown is the number density $n_{{\vec k} h}$ in same setting as before with changing interaction strengths. Thick black line is free field case with a coupling constant $y = 0$. The other lines are interacting with $y = 1$ (green line),  $y = 3$ (blue dotted line) and  $y = 5$ (red dashed line).}
\label{fig:numberdensity1}
\end{figure}

Let us next consider the case with collisions, so we have $\Gamma \neq 0$. Physically our model interaction (\ref{interaction}) corresponds to decays (and inverse decays) of the inflaton generated field $\psi$ to some lighter fermions and scalars that could be taken to be the standard model particles. Including transport equations also to these states (and the ones they couple to), and solving the whole problem with the momentum dependence, one could build an entire network of equations necessary for a realistic simulation of the particle production and thermalization at the preheating. Here our goals are more modest, and we only wish to study how the collisions affect the evolution of the particle number and coherence. As before we will set the initial chemical potential to zero so that $n_{\vec{k}h} \equiv {\bar n}_{\vec{k}h}$ and $\langle g^h_0 \rangle \equiv 1$ in the beginning. However, now this condition is not preserved, since conservation of $\langle g^h_0 \rangle$ is not respected by the collision terms. We show the evolution of the particle number and the coherence for a particular parametrization of the collision term in Fig.~\ref{fig:particle2}. Apart from including the interactions we are using the same parametrization as in the noninteracting case shown in Fig.~\ref{fig:particle0}. The difference between the interacting and noninteracting cases is quite dramatic. We see that with interactions the particle number drops between the resonance peaks, only to be regenerated again in the next resonance crossing. The drop in the particle number obviously results from the decay of the unstable particles, so that a steady flux of standard model particles (and antiparticles) is created through the resonance production and decay of the $\psi$-states. Also the growth of the coherence is damped in comparision with the noninteracting case, and both the particle number and coherence evolution settle into a stationary pattern only after a few oscillation periods. In Fig.~\ref{fig:numberdensity1} we plot the particle number evolution for varying interaction strenghts, again for our reference set of inflaton parameters. The orderly parttern of the particle number evolution over the inflaton oscillation periods remains, while the damping effect depends on the strength of the interactions in a predictable manner.

\begin{figure}
\centering
\includegraphics[angle=270,width=0.9\textwidth]{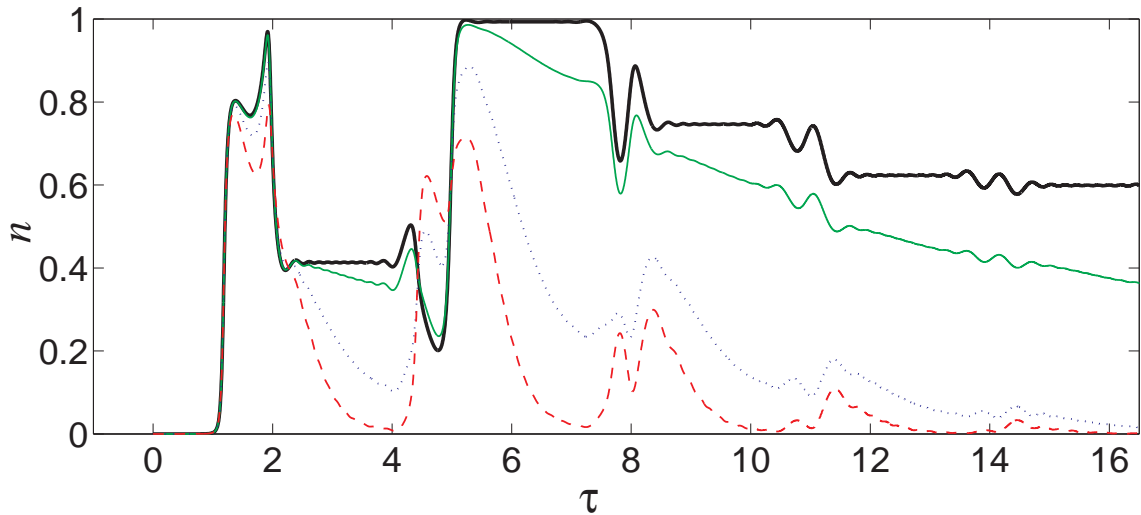}
\caption{Same as in Fig.~\ref{fig:numberdensity1}, but with the periodic mass term replaced by a damped oscillatory mass:  $ m(t) = (10 + \exp(-\gamma \tau)(15 \cos(2 \omega_{\varphi} t) + i \sin(2 \omega_{\varphi} t)))| \vec k |$, where $\gamma = 0.05$. Other parameters are again set as in all previous figures. Thick black line shows again the free field case and the interacting cases have the coupling constants  $y = 1$ (green line), $y = 3$ (blue dotted line) and $y = 5$ (red dashed line).}
\label{fig:numberdensity2}
\end{figure}

As we mentioned, getting the parametric resonance needs some fine tuning of the parameters. Moreover, when the resonance condition is not met, the evolution of the system becomes much more complicated than the orderly behaviour shown in Figs.~\ref{fig:particle0}-\ref{fig:numberdensity1}. In Fig.~\ref{fig:numberdensity2} we plot the particle number evolution for the case where the inflaton oscillation is exponentially damped, leading to a new modified mass function:
\beq
m(t) = m_0 + e^{-\gamma \tau}\Big(A \cos(2\omega_\varphi t) + i B\sin(2 \omega_\varphi t) \Big),
\label{secondmassterm}
\eeq
Although we are using a rather small damping parameter, the evolution of the particle number is dramatically changed. The orderly oscillation is replaced by an essentially chaotic evolution over the resonance crossings with changing conditions. The effect of interactions is also more pronounced, showing a very strong quantitative and qualitative dependence of the evolution on the strength of the interactions.

Our results show that making the heavy particles unstable can have dramatic effects on the pattern of the particle number production during the preheating stage of the inflation. This instability is of course a necessary requirement as these states need to decay to the standard model particles to eventually reheat the universe.  However, our formalism is well suited for a detailed study of this problem, if only the appropriate interactions are specified, the necessary transport equation network is written down also for the daughter particles, and when the inflaton dynamics is also modelled numerically along with the evolution of the transport equation network. Moreover, our methods can be used to model coherent baryogenesis~\cite{cohbaryog}, where collisions might have large quantitative effects. Let us finally note that if one introduces a CP-violation into the heavy $\psi$-decays, one can use our formalism to directly and accurately compute the evolution of the baryon (or lepton) number, in a possible decaying heavy-fermion baryogenesis scenario during the particle production at preheating.

\subsection{Decoherence and thermalization due to collisions}
\label{sec:decoherence}

\begin{figure}
\centering
\includegraphics[angle=270,width=0.9\textwidth]{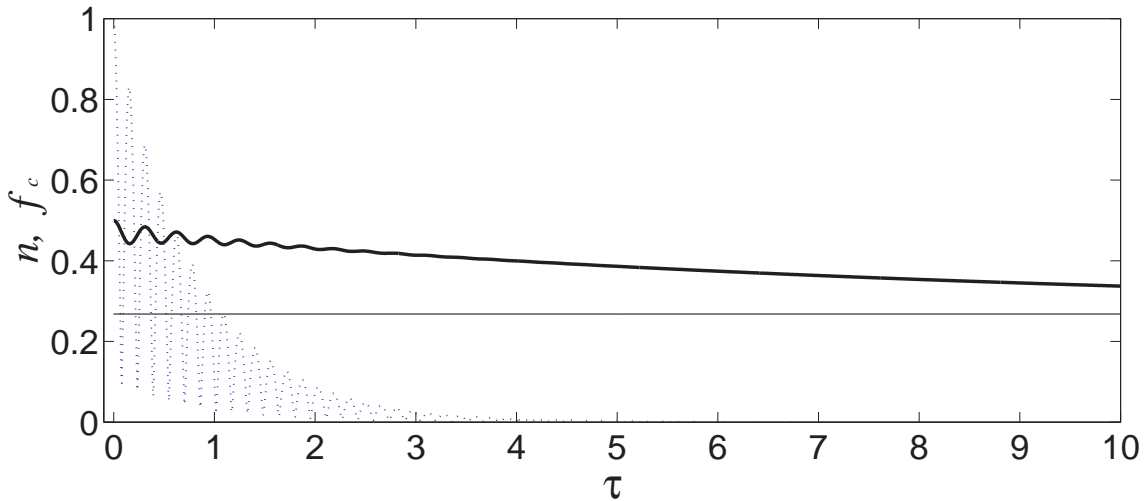}
\caption{Shown is the evolution of the number density $n_{{\vec k} h}$ (thick solid line) and the coherence function $f^h_c$ (wiggly dotted line) from an intially highly correlated out-of-equilibrium configuration towards the thermal equilibrium. The thermal equilibrium value $n_{{\rm eq}} \approx 0.27$ is shown by the thin straight line.}
\label{fig:decoherence}
\end{figure}

As our last application we consider the thermalization and decoherence of an arbitrary, correlated out-of-equilibrium density matrix  by collisions. We consider a homogenous system described by the weight function (\ref{weight2}).  We also take the mass to be a constant here, and begin with a correlated initial state at $t=0$, described by functions 
\begin{equation}
n(0) =  \bar n(0) = \frac{1}{2}, \qquad 
f_1(0) = 0 \quad {\rm and}Ê\quad f_2(0) = 1 \,.
\end{equation}
We then use the equations (\ref{rho_coll_intHOMOG}-\ref{gamma_shell}) to calculate the evolution of the density matrix. We assumed that the decay product species $q$ and $\phi$ are in thermal equilibrium and the parameters in the interaction term were taken as follows: coupling $y = 1$ and the masses $m = 10$, $m_q = 0.02$, $m_\phi = 0.1$ and the temperature $T = 10$ (all in units $|\vec k|$). The results of the calculation are shown in Fig.~\ref{fig:decoherence}. One sees clearly how the collisions smoothly damp the amplitude of the oscillating coherence on-shell functions $f_{1,2}$. As long as the coherence functions are nonzero, they affect also the mass-shell functions that show a characteristic oscillatory pattern.  After the coherence solutions are damped out, the mass-shell functions  $f_\pm$ continue to approach their equilibrium limit, which for the current parameters corresponds to $n_{\rm eq}\approx 0.27$, shown by the thin solid line in Fig.~\ref{fig:decoherence}. This behaviour can also be seen qualitatively from the Eq.~(\ref{comp_eq_constantHOMOG}) in section \ref{sec:interactions}. As a result the full 2-point Wigner function $G^<$ approaches the thermal limit $G^<_{\rm eq}$ given by Eq.~(\ref{gthermal}), as it should. The time scale for vanishing of the quantum coherence is here smaller than the thermalization time scale simply because for the parametrization chosen for the problem, the $k_0=0$-shell collision rate is much larger than the mass-shell collision rate:  $\Gamma_{00}/\Gamma_{{\rm m}} \approx 10$.

\section{Conclusions and outlook}
\label{sec:discussion}
In this paper we have developed a quantum transport formalism, simple enough to be used in practical calculations, for treating the coherent evolution of fermionic system in the presence of decohering collisions. In our derivation we used the CTP-formalism for the quantum field theory in the out-of-equilibrium conditions. The key element in the derivation was the observation that in the mean field limit and in the quasiparticle approximation in an interacting theory, the 2-point function has a singular shell structure, that in addition to the usual mass-shell solutions contain new $k_0=0$-solutions, which carry the information about the quantum coherence between particles and antiparticles. Thus the actual 2-point function can be understood as a dynamical phase space weight function, or measure, with a number of unknown on-shell coefficient functions that carry the physical information about the system. These can be interpreted as the out-of-equlibrium particle numbers on mass shells and as functions measuring quantitatively the level of quantum coherence on $k_0=0$-shell. We showed that the new coherence solutions are excluded from the spectral function by the spectral sum-rule, and that they are also eliminated from $G^<$ by the KMS-condition in the thermal equlibrium limit. The fact that coherence solutions appear only in the dynamical function makes sense, because they describe correlation between mass-shell states, and no coherence should be present without interfering physical states (Solitary coherence functions can however, in some cases describe pairs of localized virtual states~\cite{HKR1}.)

We then proceeded to show how a sensible physical density matrix can be defined by an integration procedure which involves specifying the amount of external information on the system. This information affects both the definition of the physical density matrix and its eventual equation of motion. Finally we derived an explicit dynamical equation of motion for a physical density matrix corresponding to a spatially homogenous system, where the 3-momentum and helicity are good quantum numbers.  We also computed the collision integrals in the thermal limit for a model interaction Lagrangian, showing in particular how different shells pick up different collision terms in the integration procedure, corresponding to the different external momentum configurations at different shells. Finally, we applied our formalism to the case of coherent particle production during the preheating stage of the inflation including a finite decay width to the produced heavy particles. We showed that including collisions can have dramatic effect on the evolution of the particle number, in particular when one includes a nontrivial evolution of the inflaton field. We stress that our method can be used to a quantitative analysis of the particle production during inflation if one models the inflaton oscillation and evolution numerically, and includes a network of transport equations also for the daughter states in the decay process. Finally we pointed out that the method is also suitable for a quantitative analysis of coherent baryogenesis and a computation of a baryon- or lepton number generation during decay of the preheating produced heavy states. In our last example we showed explicitly how an initially highly correlated out-of-equilibrium density matrix relaxes to a thermal equilibrium. 

We believe that the current formalism will find applications in many different contexts, also outside the cosmology. In particular we are currently extending the formalism also to the case of relativistic bosonic fields and also to nonrelativistic systems. It is also interesting to see how the usual fermionic flavour oscillation pattern arises in the present context when the formalism is extended to the case with many mixing fermion flavours~\cite{HKR1,HKRfuture}. 

\section*{Acknowledgments}
This work was partly supported by a grant from the Jenny and Antti Wihuri Foundation (Herranen) and the Magnus Ehrnrooth foundation (Rahkila).

%
%

\end{document}